\documentstyle[epsf]{article}
\newcommand{\be}{\begin{equation}}
\newcommand{\ee}{\end{equation}}
\newcommand{\bea}{\begin{eqnarray}}
\newcommand{\eea}{\end{eqnarray}}
\newcommand{\beas}{\begin{eqnarray*}}
\newcommand{\eeas}{\end{eqnarray*}}
\newcommand{\bdm}{\begin{displaymath}}
\newcommand{\edm}{\end{displaymath}}
\newcommand{\ba}{\begin{array}}
\newcommand{\ea}{\end{array}}
\newcommand{\bi}{\begin{itemize}}
\newcommand{\ei}{\end{itemize}}
\newcommand{\ben}{\begin{enumerate}}
\newcommand{\een}{\end{enumerate}}
\newcommand{\bc}{\begin{center}}
\newcommand{\ec}{\end{center}}
\newcommand{\bfl}{\begin{flushleft}}
\newcommand{\efl}{\end{flushleft}}
\newcommand{\bfr}{\begin{flushright}}
\newcommand{\efr}{\end{flushright}}
\newcommand{\bd}{\begin{description}}
\newcommand{\ed}{\end{description}}
\newcommand{\bq}{\begin{quote}}
\newcommand{\eq}{\end{quote}}
\newcommand{\bfg}{\begin{figure}}
\newcommand{\efg}{\end{figure}}
\newcommand{\bt}{\begin{table}}
\newcommand{\et}{\end{table}}
\newcommand{\btb}{\begin{tabular}}
\newcommand{\etb}{\end{tabular}}
\newcommand{\btg}{\begin{tabbing}}
\newcommand{\etg}{\end{tabbing}}

\newcommand{\kslash}
           {\mbox{$ k \hspace{-1.1ex} \mbox{/} \hspace{-0.07ex} $}}

\newcommand{\qslash}
           {\mbox{$ q \hspace{-1.1ex} \mbox{/} \hspace{-0.05ex} $}}

\begin{document}
{\small \noindent MZ-TH//96-31\\
\noindent  hep-th/9610128\\[2cm]}
\begin{center}
{\Large {\bf On knots in subdivergent diagrams\\[1cm]}}
{\large Dirk Kreimer\footnote{supported by a Heisenberg Fellowship\\email: 
kreimer@dipmza.physik.uni-mainz.de}}\\[7mm]
{\em  Dept.~of Physics}\\
{\em  Univ.~of Mainz}\\
{\em  Staudingerweg 7}\\
{\em  55099 Mainz}\\
{\em  Germany\\[2cm]}
\end{center}
\begin{abstract}
We investigate Feynman diagrams which are calculable in terms
of generalized one-loop functions, and explore how the
presence or absence of transcendentals in their counterterms
reflects the entanglement of link diagrams constructed from them
and explains unexpected relations between them.
\end{abstract}

\section{Introduction}
Recently, a connection between knot theory and renormalization theory
emerged. It initiated new results in field theory and number theory,
via the identification of knots extracted from the topology of
Feynman diagrams, with transcendentals found in their overall
divergent contributions 
\cite{habil,db,plb,pisa,bdk,bgk,bk15,BBB,4TR,BK4,db2}.
The idea explored in these papers can be described as follows.
We consider an overall divergent Feynman diagram. It is well known that after
proper renormalization of its subdivergences, its overall divergence
will depend on external parameters like masses and momenta
only in a trivial (polynomial) manner.
It further depends on the spin representation of the involved particles,
and on the topology of the diagram in a non-trivial manner. 
With the spin of the particles specified,
one can thus expect to find topological information in the values of the overall
divergences. We will now shortly summarize the results achieved so far,
and explain how the present paper fits into this context.

The simplest topologies are provided by ladder diagrams.
In \cite{habil,Bob}, the reader will find that such diagrams
provide only rational contributions in their overall divergences.
Correspondingly, these diagrams are knot-free, when mapped to link diagrams.
This mapping is achieved by consideration of the momentum
flow in the diagram \cite{habil,plb}. 
Each closed loop momentum contributes
a component of a link diagram. The mutual entanglement of these components
results from the resolution of vertices into over- and
undercrossings, and from the topology of the diagram.

Using this approach, it is shown
in \cite{habil} that the transcendentals
$\zeta(2l-3)$ (odd zetas) appear in $l$-loop Feynman diagrams which deliver 
$(2,2l-3)$ torus knots. These are the simplest positive knots,
and they are generated by Feynman diagrams which are distinguished from the ladder
topology by only one propagator, which crosses all the other rungs.
 
Up to the five-loop level, only these knots appear, and correspondingly,
all transcendentals in counter\-terms are of odd zeta type.
In \cite{plb}, it is demonstrated that a certain six-loop graph
generates a different knot, with
eight crossings. This is a success for knot theory: precisely
such graphs deliver for the first time a new transcendental,
a double sum of weight eight, which was first observed by David Broadhurst
\cite{1985}.

Such multiple sums are known in the literature as Euler/Zagier sums
\cite{db,BBB,db2,Zag}. The restricted class of non-alternating
sums is known as multiple zeta values (MZVs) \cite{db2}.

In \cite{pisa}, these lines of thinking were extended to the seven-loop
level, and the scheme independent contribution to the seven-loop
$\beta$-function of $\phi^4$ theory was calculated.
This involved only diagrams free of subdivergences.
Again, the match between knots and transcendentals was striking,
and culminated in the identification of the first irreducible
triple sum of weight eleven with the unique positive four-braid knot.

In \cite{db}, Broadhurst enumerates alternating sums (Euler/Zagier sums),
inspired by knot and field theory. The results inform number theory,
by answering the question how many of such transcendentals are independent
over the rational numbers. Further results appeared in \cite{BBB}.

In \cite{bgk}, we use the $\epsilon$-expansion of critical exponents
to enlargen our knot-to-number dictionary.
This implicitly incorporates graphs with subdivergences via large $N$ methods.
 
In \cite{bk15}, we identify whole classes of positive knots
with MZVs up to 15 crossings, and conjecture
the enumeration of irreducible MZVs. This conjecture is further
confirmed in \cite{db2}.

In \cite{4TR}, an explanation for the connection between knot theory and
renormalization theory is given, by indicating how the overall divergences
of Feynman diagrams (restricted to diagrams free of subdivergences)
provide a weight system, and thus are related to knot invariants of finite
type. The investigations are restricted to the study of graphs without
subdivergences.

In \cite{BK4}, examples at the four-and five-loop level
confirm these findings.

When one carefully studies the above papers, one notices that
they barely discuss graphs with subdivergences. In this paper,
we want to close the gap.
It is the purpose of this paper to compare the UV-divergent behaviour
of Feynman graphs which have subdivergences
with the behaviour of link and knot diagrams
associated to these diagrams. 

In the before-mentioned papers the topologies of the diagrams were
complicated, but in most cases free of subdivergences. Now, we focus on
fairly simple topologies with subdivergences.
From \cite{habil}, we know one result concerning
graphs with subdivergences: that simple topologies, whose
forest structure is strictly nested or purely
overlapping, provide only rational
numbers in their overall divergences. These results
are confirmed by independent methods in \cite{Bob}. 
Here, we generalize such cases
by dressing internal propagators with
simple rainbow diagrams which generate  disjoint subdivergences.

We
consider in this paper Feynman graphs which are calculable in terms of
$G$-functions \cite{chet}. 
As a specific example we consider vertex
functions at zero momentum transfer (zmt) in Yukawa theory,
with fermions coupling to a massless scalar field.
We are thus restricted to the study of nested divergences
as a starting point. 
From the results in \cite{habil}, and from
the general structure of renormalization theory, we know that 
the entanglement of subdivergences for the overlapping case
follows a similar pattern, only that one has to sum over all
maximal forests of these overlapping divergences. 
Thus, the combinations of $G$-functions generated by overlapping
divergences is ultimately a sum of the combinations which
arise from nested divergences, in full accord with the
forest formula \cite{Zimm}. This was the motivation to choose
vertex-functions as an appropriate testing ground to explore
the connection between renormalization and link diagrams in the
context of graphs with subdivergences.

The sole purpose of these vertex functions
is to serve as a visualization for the 
combinations of $G$-functions 
which we want to consider. The only transcendentals
which appear in $G$-functions stem from the
$\zeta$-function evaluated at integer argument
\be
\Gamma(1-z)=e^{\gamma z}
\exp{(\sum_{j=2}^\infty \frac{\zeta(j) z^j}{j})}.\label{eze}
\ee 
The only
knots we expect
to see are $(2,q)$ torus knots, according to the identification
of $\zeta(q)$ with these knots in \cite{habil,plb,pisa}.

\section{Definitions}
\begin{figure}
\begin{center}
\epsfysize=5cm 
\fbox{
\epsfbox{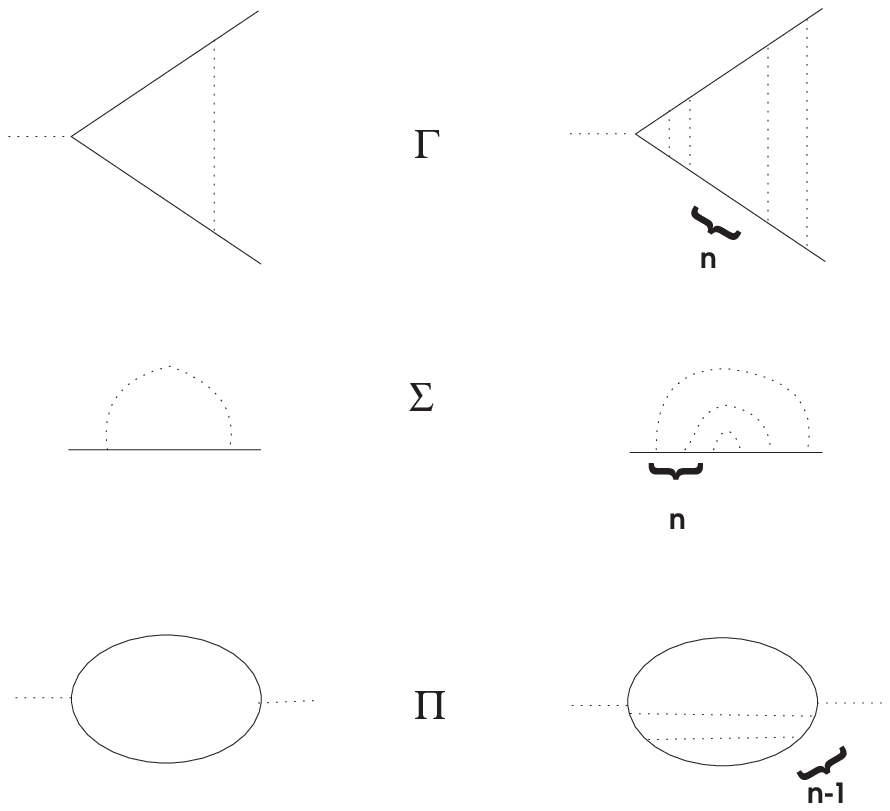}}
\caption{Basic definitions for one-loop functions 
$\Gamma^{[1]},\Sigma^{[1]},\Pi^{[1]}$, on the lhs, and
their generalization to $n$-loop functions 
$\Gamma^{[n]},\Sigma^{[n]},\Pi^{[n]}$,
on the rhs. Dashed lines indicate the massless  
spin-0 boson, while
the solid line represents a massless fermion.}
\label{gfc1}
\end{center}
\end{figure}

Our input is a set of generalized one-loop functions.
We define\footnote{We work in
the $\overline{MS}$-scheme, and thus omit 
irrelevant factors of $\log(4\pi)$ and
$\gamma_E$ throughout the paper.}
\begin{eqnarray}
G(j_1,j_2)[q^2] & \equiv & 
\int \frac{d^Dk}{N}\frac{1}{[(k+q)^2]^{1+j_1\epsilon}
[k^2]^{1+j_2\epsilon}}\nonumber\\
 & := &
[q^2]^{-(j_1+j_2+1)\epsilon}\times\label{e1}\\
 &  &  
\times\frac{\Gamma[1+(j_1+j_2+1)\epsilon]\Gamma[1-(j_1+1)\epsilon]
\Gamma[(1-(j_2+1)\epsilon)]}{
((j_1+j_2+1)\epsilon)\Gamma[(2-(j_1+j_2+2)\epsilon)]
\Gamma[(1+j_1\epsilon)]\Gamma[(1+j_2\epsilon)]},\nonumber
\end{eqnarray}
where $D=4-2\epsilon$ and the last equation fixes the
normalization.
In this notation, the one-loop radiative corrections
of the vertex, the fermion and the scalar propagator in massless
Yukawa theory become
\begin{eqnarray}
\Gamma^{[1]}(q) & = & G(0,0)[q^2],\label{eg1}\\
\Sigma^{[1]}(q) & = & -\frac{1}{2}\qslash G(0,0)[q^2],\label{es1}\\
\Pi^{[1]}(q)    & = & 2q^2G(0,0)[q^2],\label{ep1}
\end{eqnarray}
for the (one-loop) vertex-correction at zmt, $\Gamma^{[1]}$; the fermion
self-energy, $\Sigma^{[1]}$; and the self-energy of the scalar boson,
$\Pi^{[1]}$.
These corrections refer to the graphs given in Fig.(\ref{gfc1})
on the lhs.
We can easily generalize this to $n$-loop ladder corrections at the vertex
$\Gamma^{[n]}$,
and to $n$-loop rainbow corrections for the self-energy of the
fermion propagator, $\Sigma^{[n]}$.
For the scalar boson, we restrict ourselves to dressings of
one internal fermion line, which gives us a function
$\Pi^{[n]}$. 
In Fig.(\ref{gfc1}) we define these
functions diagrammatically on the rhs, and analytically
in Eqs.(\ref{egn},\ref{esn},\ref{epn}) below.

We denote such diagrams as simple. Such simple topologies
deliver rational counterterms, after we absorb the subdivergences
by counterterms \cite{habil, Bob}. In short, their overall
divergence provides a Laurent series in $(D-4)$ which has
coefficients in ${\bf Q}$, the rational numbers.

Now these functions are the building blocks for the
Feynman diagrams to be considered in this paper. 
According to the results in 
\cite{habil}
we expect a connection to knot theory to appear after renormalization of
subdivergences. This is a natural constraint. When the
dependence on external parameters is removed via renormalization of 
subdivergences we shall expect to find an overall divergence which solely
reflects the topology of the diagram under consideration.

The functions defined in Fig.(1) can be defined 
analytically by the following iterations
\bea
\Gamma^{[n]}(q) & := & \int d^Dk \;\Gamma^{[n-1]}(k)\frac{1}{\kslash}
\frac{1}{\kslash}\frac{1}{(k+q)^2},\label{egn}\\
\Sigma^{[n]}(q) & := & \int d^Dk \;\frac{1}{\kslash}\Sigma^{[n-1]}(k)
\frac{1}{\kslash}\frac{1}{(k+q)^2},\label{esn}\\
\Pi^{[n]}(q) & := & Tr\left(\int d^Dk\; \frac{1}{\kslash}\Sigma^{[n-1]}(k)
\frac{1}{\kslash}\frac{1}{\kslash+\qslash}\right),\label{epn}\\
 & & \Gamma^{[0]}(k):=1,\;\; \Sigma^{[0]}(k):=\kslash. 
\eea
Using 
\be
P_n\equiv P_n(\epsilon):=\prod_{i:=0}^{n-1}G(0,i)[q^2=1],\label{epp}
\ee
one obtains by a standard calculation the following
explicit expressions
\bea
\Gamma^{[n]}(q) & = & [q^2]^{-n\epsilon}P_n,\label{ege}\\
\Sigma^{[n]}(q) & = & (-\frac{1}{2})^n\qslash 
[q^2]^{-n\epsilon}P_n,\label{ese}\\
\Pi^{[n]}(q) & = & 2(-\frac{1}{2})^{n-1} q^2 
[q^2]^{-n\epsilon}P_n.\label{epe}
\eea
These Green functions are unrenormalized.
For $n>1$, they contain subdivergences.
To find their overall counterterms we have to renormalize 
these subdivergences 
first.
For $n>1$ we incorporate
the subtraction of subdivergences as follows:
\bea
\Gamma^{[n]} & \rightarrow &
\bar{\Gamma}^{[n]} := 
\Gamma^{[n]}-\sum_{i:=1}^{n-1}Z_\Gamma^{[i]}\Gamma^{[n-i]},\\
 & & Z_\Gamma^{[n]} := < \bar{\Gamma}^{[n]}>,\; 
Z_\Gamma^{[1]}\equiv 
<\Gamma^{[1]}>,\\
\Sigma^{[n]} & \rightarrow &
\bar{\Sigma}^{[n]} := 
\Sigma^{[n]}-\sum_{i:=1}^{n-1}Z_\Sigma^{[i]}\Sigma^{[n-i]},\\
 & & Z_\Sigma^{[n]} := < \bar{\Sigma}^{[n]}/\qslash>,\; 
Z_\Sigma^{[1]}\equiv 
<\Sigma^{[1]}/\qslash>,\\ 
\Pi^{[n]} & \rightarrow &
\bar{\Pi}^{[n]} := 
\Pi^{[n]}-\sum_{i:=1}^{n-1}Z_\Sigma^{[i]}\Pi^{[n-i]},\\
 & & Z_\Pi^{[n]} := < \bar{\Pi}^{[n]}/q^2>,\; Z_\Pi^{[1]}\equiv 
<\Pi^{[1]}/q^2>.
\eea
In angle brackets $<\ldots>$ we project
onto the proper pole part of the Laurent series in $\epsilon$ and
evaluate at $q^2=1$. The
renormalization of subdivergences
is achieved in the $\overline{MS}$-scheme. We introduced 
renormalization $Z$-factors 
$Z_{\Gamma,\Sigma,\Pi}^{[n]}$ taylored for our purposes.
They remove the overall divergence at the indicated loop order
in our simple Green functions.
These $Z$-factores are totally expressible in terms
of the $P_n$ functions, for example
\be
Z^{[3]}_\Gamma = <P_3-\overbrace{<P_1>}^{Z_\Gamma^{[1]}}P_2
-\overbrace{<P_2-<P_1>P_1>}^{Z_\Gamma^{[2]}}P_1>.
\ee
Renormalized Green functions (usually referred to by
bold letters) are obtained by subtracting the
remaining overall divergences
\bea
{\bf \Upsilon}^{[n]} & := & \bar{\Upsilon}^{[n]}-<\bar{\Upsilon}^{[n]}>,
\label{e2}\\
 & & \Upsilon \in \{\Gamma,\Sigma,\Pi \}.\nonumber
\eea

To go from self-energies to propagators, we use
\begin{eqnarray}
\Delta_F & := & \frac{i}{q^2-\Pi},\\
S_F & := & \frac{i}{\qslash-\Sigma},\\
\Pi & := & \sum_{n:=1}^\infty \Pi^{[n]},\\
\Sigma & := & \sum_{n:=1}^\infty \Sigma^{[n]}.
\end{eqnarray}
Quite often we only want to use a terminating
geometric series for self-energies
with rainbows of a fixed loop number $i$, expanded
to a series of degree $k$, so that we define
\bea
\Delta_F^{[i,k]}(q) & := & \frac{i}{q^2}
\left[\frac{\Pi^{[i]}(q)}{
q^2}\right]^k,\label{esik}\\
S_F^{[i,k]}(q) & := & \frac{i}{\qslash}\left[\frac{
\Sigma^{[i]}(q)}{
\qslash}\right]^k.\label{epik}
\eea
The corresponding renormalized quantities 
${\bf \Delta_F}^{[i,k]}$
and
${\bf S_F}^{[i,k]}$
are obtained by replacing the self-energies $\Sigma^{[i]},\Pi^{[i]}$
by the corresponding renormalized 
${\bf \Sigma}^{[i]},{\bf \Pi}^{[i]}$.

\begin{figure}
\begin{center}
\epsfysize=4cm \fbox{\epsfbox{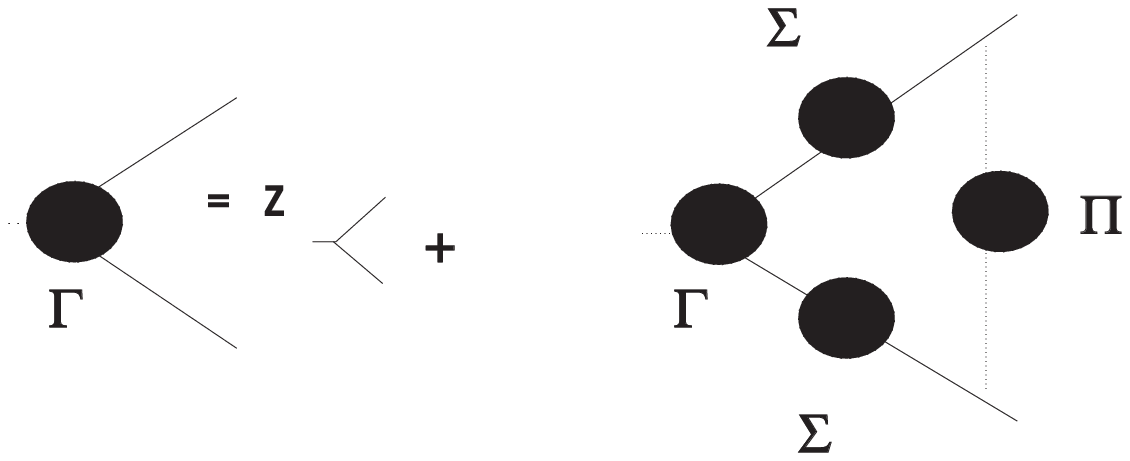}}
\caption{Feynman diagrams are generated by the Schwinger-Dyson
equation for the vertex as indicated in the figure.
We allow for the selfenergy insertions $\Sigma,\Pi$
defined in the text.}
\label{gfc2}
\end{center}
\end{figure}

Having defined all these quantities, we now
consider the diagrams of Fig.(\ref{gfc2}). 
Their generic structure is
\be
\Gamma\sim \int [S_F^2 \Delta_F]^r.
\ee

They come from dressing internal propagators and iterating
the vertex as given by the reduced Schwinger Dyson equation 
of Fig.(\ref{gfc2}). 
Propagator dressed with disjoint subdivergences
are obtained from expanding
the one-particle irreducible functions defined above
in a power series in the selfenergies, using Eqs.(\ref{esik},\ref{epik}).

We investigate the contribution of the so-iterated vertex
to the MS $Z$-factor. Especially, we are interested to what
extent transcendentals remain after the renormalization of subdivergences.
So, in contrast to the simple case, we expect coefficients
to be $\not\in {\bf Q}$. 
Especially, we expect to see
the transcendentals
$\zeta(2l+1)$, generated from Eq.(\ref{eze}) and
corresponding to the $(2,2l+1)$ torus knots,
familiar from previous results \cite{habil,plb,pisa}.
We once more stress that for the case of  simple topologies,
with undressed internal propagators as in Fig.(\ref{gfc1}) on the rhs,
all transcendentals vanish after the renormalization of
proper subdivergences. 

\begin{figure}
\begin{center}
\epsfysize=4cm \fbox{\epsfbox{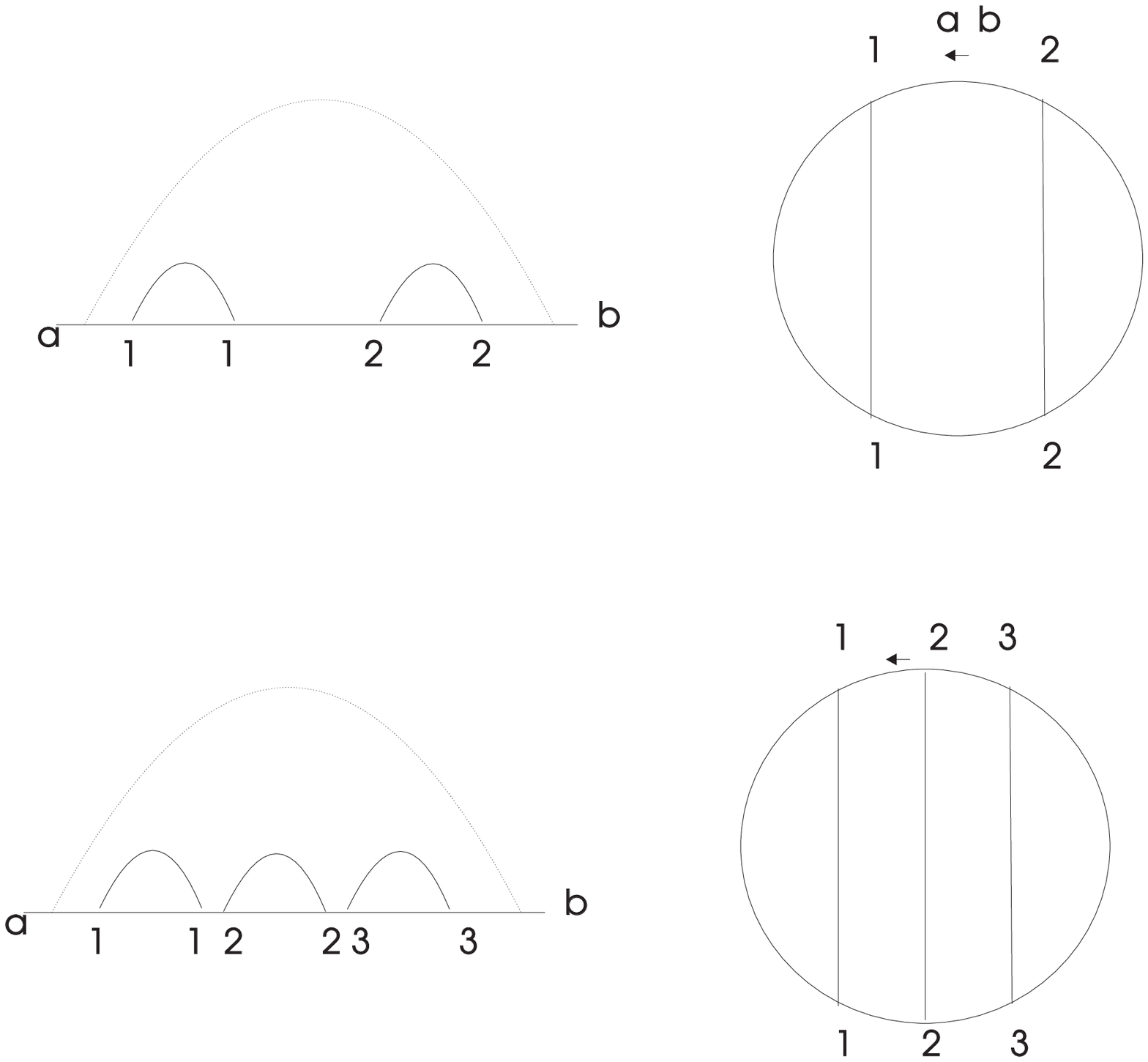}}
\caption{The most basic dressings are topologically equivalent
to ladder diagrams. Three selfenergy insertions are needed to get a topology
different from a ladder topology. The topology can be encoded as a (Gauss-)code,
by reading off sequences like $\{1,1,2,2,3,3\}$ for the case
of three disjoint subdivergences, and $\{1,2,3,3,2,1\}$ in the four-loop
ladder case.}
\label{gfc0n}
\end{center}
\end{figure}

The most basic example appears when we start to dress $\Gamma^{[1]}$
with one-loop self-energies $\Sigma^{[1]}$ or $\Pi^{[1]}$.
As long as there are only one or two such dressings, the topology
of the resulting Feynman diagrams is still similar to a ladder
topology, as Fig.(\ref{gfc0n}) exhibits. We thus expect trancendentals to cancel,
as we claim that there appearance reflects a change from the
ladder topology.
Indeed, the next section
will reveal that transcendentals only remain
in case of more than two disjoint subdivergences.
\section{An elementary example}
We start with the consideration of
Feynman graphs before the subtraction of subdivergences,
with $n$ one-loop bubbles at the fermion line and
$m$ one-loop insertions at the boson line,
for different $n$ and $m$, but with $n+m=l$ fixed. 

To compare cases with a varying number of fermion-
or boson self-energies, from now on
we normalize Green functions by removing the factors of 
$2$ and $(-1/2)$ in Eqs.(\ref{ese},\ref{epe}) for the selfenergies.
In so doing, we employ the fact that 
$Z_\Sigma^{[n]}$ and $Z_\Pi^{[n]}$
are related by such simple factors, according to Eqs.(\ref{ese},\ref{epe}).
After such a normalization (which removes the burden to list
superflous factors of 2 in our tables) we have $Z_\Gamma^{[n]}=
Z_\Sigma^{[n]}=Z_\Pi^{[n]}=:Z^{[n]}$. 
This normalization is included in all
the results in the tables and adopted in all what follows.

\begin{figure}
\begin{center}
\epsfysize=3cm \fbox{\epsfbox{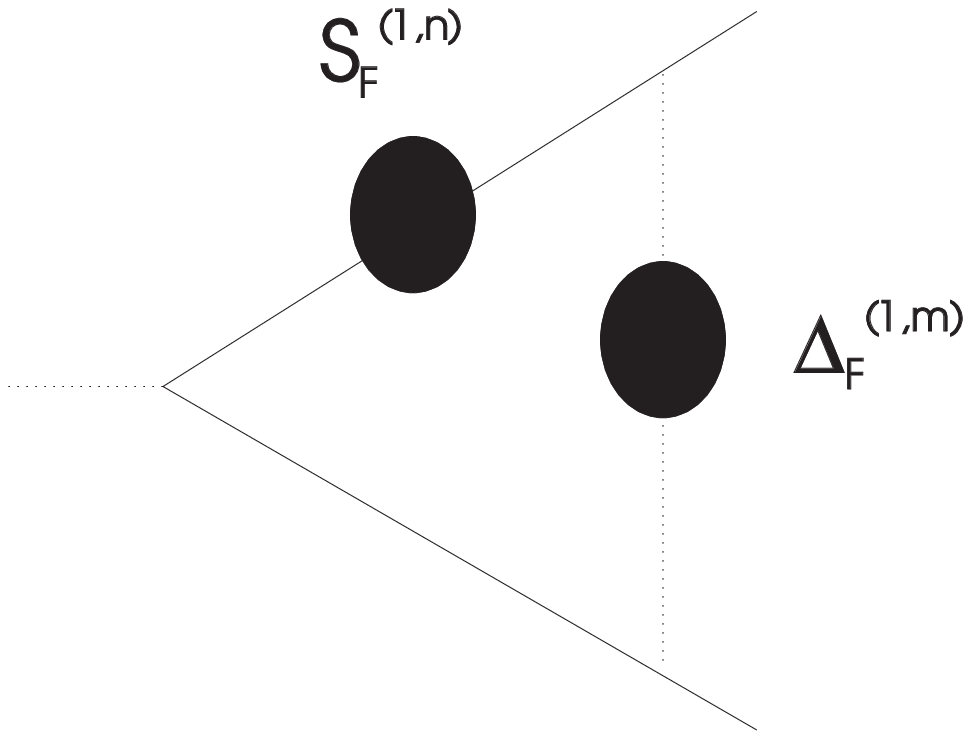}}
\caption{$\bar{\Gamma}(n,m)$.}
\label{gfc3}
\end{center}
\end{figure}

We will consider a function $\Gamma(n,m)$ defined as
\bea
\Gamma(n,m)(q) & := & 
\int d^Dk \frac{1}{\kslash}
\;S_F^{[1,n]}(k)
\;\Delta_F^{[1,m]}(k+q)\nonumber\\
 & = & 
[G(0,0)]^{n+m}\;G(0,n+m)[q^2]^{-(n+m+1)\epsilon}.
\eea
Renormalizing the subdivergences delivers
\bea
\bar{\Gamma}(n,m)(q) & := & 
\int d^Dk \frac{1}{\kslash}
\;{\bf S_F}^{[1,n]}(k)
\;{\bf \Delta_F}^{[1,m]}(k+q)\\
& = &  
\sum_{i=0}^{n+m}{n+m \choose i}
G(0,0)^{i}[-<G(0,0)>]^{n+m-i}G(0,i)[q^2]^{-(i+1)\epsilon}.
\eea
The corresponding Feynman graphs are  given in Fig.(\ref{gfc3}).
We easily evaluate them using Eqs.(\ref{ese},\ref{epe}). 
We are not so much
interested in an all order result, but rather compare the
results before and after subtraction of subdivergences.

Table(1)
shows that the graphs in general differ for different $n,m$.
A few words about the conventions in the tables are in order.
Only divergent parts of Green functions are given.
The notation $\bar{\Gamma}$ refers to Green functions with subtracted
subdivergences. Thus, all the dependence on $q^2$ has dropped out.
Only in Table(1) we give for comparison a few cases for unsubtracted
Green-functions, in the first six entries. For these entries,
we have set $q^2=1$.
\begin{table}
\rule{\textwidth}{.5mm}\\
$\Gamma(2,0) :=  
\;\epsilon^{(-1)}( - 1/2\;\zeta(2)   + 44/3) + 
8/3\;\epsilon^{(-2)}
 + 1/3 \;\epsilon^{(-3)}$\\
\rule{\textwidth}{.5mm}\\
$\Gamma(1,1) :=  \;\epsilon^{(-1)}( - 1/2\;\zeta(2)  + 
44/3) +  
8/3\;\epsilon^{(-2)}
 + 1/3 \;\epsilon^{(-3)}$\\
\rule{\textwidth}{.5mm}\\
$\Gamma(3,0) :=  \;\epsilon^{(-1)}
( - 59/6\;\zeta(3)  - 11/2\;\zeta(2) + 475/4) +  
\;\epsilon^{(-2)}( - 1/2\;\zeta(2)  + 79/4) +  
11/4\;\epsilon^{(-3)} + 1/4 
\;\epsilon^{(-4)}$\\
\rule{\textwidth}{.5mm}\\
$\Gamma(2,1) :=  \;\epsilon^{(-1)}( - 83/6\;\zeta(3) -
11/2\;\zeta(2) + 475/4) +  
\;\epsilon^{(-2)}( - 1/2\;\zeta(2) + 79/4) + 
11/4\;\epsilon^{(-3)} + 1/4 
\;\epsilon^{(-4)}$\\
\rule{\textwidth}{.5mm}\\
$\Gamma(5,1) := 
\;\epsilon^{(-1)} ( - 113767/42 \;\zeta(6) - 155308/7 \;\zeta(5) + 1087/8 \;\zeta(4) 
\;\zeta(2)
 - 66307/7 \;\zeta(4) + 299209/126 \;\zeta(3)^{2} + 
5470/3 \;\zeta(3) \;\zeta(2) - 437600/7 
\;\zeta(3) - 49/48 \;\zeta(2)^{3} + 427/2 \;\zeta(2)^{2} 
- 10608 \;\zeta(2) + 1451904/7) + 
\;\epsilon^{(-2)} (
 - 38827/35 \;\zeta(5) - 5435/7 \;\zeta(4) + 547/6 \;\zeta(3) \;\zeta(2) - 133468/21 
\;\zeta(3)
 + 35/2 \;\zeta(2)^{2} - 1200 \;\zeta(2) + 177792/7) + 
\;\epsilon^{(-3)} ( - 1087/28 \;\zeta(4) - 
10940/21 \;\zeta(3) + 7/8 \;\zeta(2)^{2} - 122 \;\zeta(2) + 21216/7) + \;\epsilon^{(-4)}
 ( 
- 547/21 
\;\zeta(3) - 10 \;\zeta(2) + 2400/7) + \;\epsilon^{(-5)} ( - 1/2 \;\zeta(2) + 244/7) + 
20/7 \;\epsilon^{(-6)}
 + 1/7 \;\epsilon^{(-7)}$\\
\rule{\textwidth}{.5mm}
$\Gamma(4,2) := \;\epsilon^{(-1)} ( 
- 132667/42 \;\zeta(6) - 181348/7 \;\zeta(5) + 1339/8 \;\zeta(4) 
\;\zeta(2)
 - 81679/7 \;\zeta(4) + 452929/126 \;\zeta(3)^{2} + 
6730/3 \;\zeta(3) \;\zeta(2) - 538400/7 
\;\zeta(3) - 49/48 \;\zeta(2)^{3} + 427/2 \;\zeta(2)^{2} -
 10608 \;\zeta(2) + 1451904/7) + 
\;\epsilon^{(-2)} 
( - 45337/35 \;\zeta(5) - 6695/7 \;\zeta(4) + 673/6 \;\zeta(3) \;\zeta(2) - 164212/21 
\;\zeta(3)
 + 35/2 \;\zeta(2)^{2} - 1200 \;\zeta(2) + 177792/7) + \;\epsilon^{(-3)}
 ( - 1339/28 \;\zeta(4) - 
13460/21 \;\zeta(3) + 7/8 \;\zeta(2)^{2}- 
122 \;\zeta(2) + 21216/7) + \;\epsilon^{(-4)} ( - 
673/21 
\;\zeta(3) - 10 \;\zeta(2) + 2400/7) + \;\epsilon^{(-5)}
 ( - 1/2 \;\zeta(2) + 244/7) + 20/7 \;\epsilon^{(-6)}
 + 1/7 \;\epsilon^{(-7)}$\\
\rule{\textwidth}{.5mm}
$\overline{\Gamma}(1) := 1/2 \;\epsilon^{(-1)} - 1/2 \;\epsilon^{(
-2)}$\\
\rule{\textwidth}{.5mm}\\
$\overline{\Gamma}(2) :=  - 1/3 \;\epsilon^{(-1)} - 1/3 \;\epsilon^{(-2)} + 1/3 
\;\epsilon^{(-3)}$\\
\rule{\textwidth}{.5mm}\\
$\overline{\Gamma}(3) :=  
\;\epsilon^{(-1)}( - 1/2\;\zeta(3) + 1/4) + 1/4 \;\epsilon^{(-2)} + 1/4 
\;\epsilon^{(-3)} -
 1/4 \;\epsilon^{(-4)}$\\
\rule{\textwidth}{.5mm}\\
$\overline{\Gamma}(4) :=  
\;\epsilon^{(-1)}(3/5\;\zeta(4) - 2/5\;\zeta(3) - 1/5) +  
\;\epsilon^{(-2)}(2/5\;\zeta(
3) - 1/5) - 
1/5 \;\epsilon^{(-3)} - 1/5 \;\epsilon^{(-4)} + 1/5 \;\epsilon^{(-5)}$\\
\rule{\textwidth}{.5mm}\\
$\overline{\Gamma}(5) :=  
\;\epsilon^{(-1)}( - \;\zeta(5) + 1/2\;\zeta(4) + 1/3\;\zeta(3) + 1/6) +  
\;\epsilon^{(-2)}
( - 1/2\;\zeta(4) + 1/3\;\zeta(3) + 1/6) +  \;\epsilon^{(-3)}( - 1/3\;\zeta(3) + 1/6
) + 1/6 \;\epsilon^{(-4)} + 1/6 \;\epsilon^{(-5)} - 1/6 \;\epsilon^{(-6)}$\\
\rule{\textwidth}{.5mm}\\
$\overline{\Gamma}(7) := \;\epsilon^{(-1)} ( - 9/4 \;\zeta(7) + 5/4 \;\zeta(6) + 3/4 \;\zeta(5) - 
3/4 \;\zeta(4) \;\zeta(3) +
 3/8 \;\zeta(4) + 1/4 \;\zeta(3)^{2} + 1/4 \;\zeta(3) + 1/8) + \;\epsilon^{(-2)} ( - 5/4 \;\zeta(6) 
+ 3/4 \;\zeta(5) 
+ 3/8 \;\zeta(4) - 1/4 \;\zeta(3)^{2} + 1/4 \;\zeta(3) + 1/8) + \;\epsilon^{(-3)} ( - 3/4 
\;\zeta(5) + 3/8 \;\zeta(4) + 1/4 \;\zeta(3) + 1/8) + \;\epsilon^{(-4)} ( - 3/8 \;\zeta(4) + 
1/4 \;\zeta(3)
 + 1/8) + \;\epsilon^{(-5)} ( - 1/4 \;\zeta(3) + 1/8) + 1/8 \;\epsilon^{(-6)} + 1/8 \;\epsilon^{(-7)} - 
1/8 \;\epsilon^{(-8)}$\\
\rule{\textwidth}{.5mm}
$\overline{\Gamma}(10) := 
\;\epsilon^{(-1)} (102/11 \;\zeta(10) - 170/33 \;\zeta(9) - 63/22 \;\zeta(8) + 36/11 
\;\zeta(7)
 \;\zeta(3) - 18/11 \;\zeta(7) + 30/11 \;\zeta(6) \;\zeta(4) - 20/11 \;\zeta(6) \;\zeta(3) - 
10/11 
\;\zeta(6) + 18/11 \;\zeta(5)^{2} - 18/11 \;\zeta(5) \;\zeta(4) - 
12/11 \;\zeta(5) \;\zeta(3) - 6/11 
\;\zeta(5) - 9/22 \;\zeta(4)^{2} + 6/11 \;\zeta(4) \;\zeta(3)^{2} - 
6/11 \;\zeta(4) \;\zeta(3) - 3/11 
\;\zeta(4) - 4/33 \;\zeta(3)^{3} - 2/11 \;\zeta(3)^{2} - 2/11 \;\zeta(3) - 1/11) + \;\epsilon^{(-2)}
 (170/33 \;\zeta(9) - 
63/22 \;\zeta(8) - 18/11 \;\zeta(7) + 20/11 \;\zeta(6) \;\zeta(3) - 10/11 \;\zeta(6)
 + 18/11 \;\zeta(5) \;\zeta(4) - 12/11 \;\zeta(5) \;\zeta(3) - 6/11 \;\zeta(5) - 9/22 
\;\zeta(4)^{2} - 
6/11 \;\zeta(4) \;\zeta(3) - 3/11 \;\zeta(4) + 4/33 \;\zeta(3)^{3} - 
2/11 \;\zeta(3)^{2} - 2/11 
\;\zeta(3) - 1/11) + \;\epsilon^{(-3)} (
63/22 \;\zeta(8) - 18/11 \;\zeta(7) - 10/11 \;\zeta(6) + 
12/11 
\;\zeta(5) \;\zeta(3) - 6/11 \;\zeta(5) + 9/22 \;\zeta(4)^{2} - 
6/11 \;\zeta(4) \;\zeta(3) - 
3/11 \;\zeta(4) - 
2/11 \;\zeta(3)^{2} - 2/11 \;\zeta(3) - 1/11) + \;\epsilon^{(-4)} (18/11 \;\zeta(7) - 10/11 
\;\zeta(6) - 
6/11 \;\zeta(5) + 6/11 \;\zeta(4) \;\zeta(3) - 3/11 \;\zeta(4) - 2/11 \;\zeta(3)^{2} - 
2/11 \;\zeta(3) - 1/11) + \;\epsilon^{(-5)} (10/11 \;\zeta(6) - 6/11 \;\zeta(5) - 3/11 \;\zeta(4) 
+ 2/11 
\;\zeta(3)^{2} - 2/11 \;\zeta(3) - 1/11) + \;\epsilon^{(-6)} (6/11 \;\zeta(5) - 3/11 \;\zeta(4) - 
2/11 
\;\zeta(3) - 1/11) + \;\epsilon^{(-7)} (3/11 \;\zeta(4) - 2/11 \;\zeta(3) - 1/11) + 
\;\epsilon^{(-8)} (2/11 
\;\zeta(3) - 1/11) - 1/11 \;\epsilon^{(-9)} - 1/11 \;\epsilon^{(-10)} + 1/11
 \;\epsilon^{(-11)}$\\
\rule{\textwidth}{.5mm}

\caption{$<\Gamma(n,m)>$ for various
values of $(n_1,n_2)$. After the renormalization
of subdivergences we find that the results for $<\overline{\Gamma}(n,m)>$
depend only on the sum $l=n+m$, which we give as $<\overline{\Gamma}(l)>$.
The transcendental $\zeta(l)$ appears at $l+1$ loops.}
\label{t1}
\end{table}
Now let us consider Table(1). 
While still $\Gamma(1,1)=\Gamma(2,0)$, we have
$\Gamma(2,1)\not=\Gamma(3,0)$ and
$\Gamma(5,1)\not=\Gamma(4,2)$, cf.~Table(\ref{t1}).\footnote{In all
the tables, expressions like $a/b\;\epsilon^{(-r)}$ shall be read as
$\frac{a}{b}\frac{1}{\epsilon^r}$.}
Further calculations confirm that in general
$\Gamma(n,m)\not=\Gamma(n^\prime,m^\prime)$ for $n+m\geq 3$, where
we always have $n+m=n^\prime+m^\prime$.

Now consider the graph after subtraction of subdivergences.
A symmetry is emerging. The results only depend
on the sum $l=n+m$. Table(\ref{t1}) shows results up to the eleven loop
level, given as $\overline{\Gamma}(l)$. The pattern indicated
in the table continues to higher loop levels, as checked by systematic
calculations. 
Note that the results in Table(\ref{t1}) are given 
in a normalization which makes the observed symmetry most obvious. 
In this normalization we simply
have dropped irrelevant factors of 2 or $1/2$ which come
from the calculation of selfenergies 
according to Eqs.(\ref{ese},\ref{epe},\ref{esik},\ref{epik}).
As a result our
Feynman diagrams now solely indicate the nesting of
$G$-functions. With this normalization, all the basic one-loop
functions have the same divergence. 

With these conventions in place,
the explicit formulae for
$<\overline{\Gamma}(2,1)>$ = $<\overline{\Gamma}(3,0)>$ are as follows:
\begin{eqnarray}
<\overline{\Gamma}(2,1)> & = & <G(0,0)^3\;G(2,1)-
G(0,0)^2\;G(2,0)
Z^{[1]}\nonumber\\
 & & -2G(0,0)^2\;G(1,1)Z^{[1]}
 +3G(0,0)\;G(1,0)[Z^{[1]}]^2\nonumber\\
 & &-G(0,0)[Z^{[1]}]^3>,\\
<\overline{\Gamma}(3,0)> & = & <G(0,0)^3\;G(3,0)-3G(0,0)^2\;G(2,0)Z^{[1]}
 +3G(0,0)\;G(1,0)[Z^{[1]}]^2\nonumber\\
 & & -G(0,0)[Z^{[1]}]^3>,
\end{eqnarray}
where we used $G(1,0)=G(0,1)$. 
In both equations, the first term on the rhs is
the bare Green function and the other terms renormalize
the subdivergences.

As in \cite{habil}, we now encode the momentum flow of the diagrams
into link diagrams, associated to the Feynman diagrams.
In this transition, the underlying one-loop vertex-function
is simply represented by a circle, 
and all information as to where external particles couple is 
lost. Thus, the observed symmetry is obvious, as we clearly 
see in Fig.(\ref{gfc0n}).

In fact, we can also prove this symmetry from basic properties of 
our Green functions. We will modify the propagator $\Delta_F$ in a way
which will not change its high energy behaviour.
We use\footnote{And a similar relation for $S_F$, if necessary.} 
\be
\Delta_F^{[1,r]}(k)=\Delta_F^{[1,r-j]}(k)\;(-ik^2)\; 
\Delta_F^{[1,j]}(k),\;\;\forall j\label{efac}
\ee
and define 
\be
\nabla_F^{[r,j]}(k,q)=\Delta_F^{[1,r]}(k+q)-
\Delta_F^{[1,r-j]}(k+q)\;(-ik^2)\;\Delta_F^{[1,j]}(k),
\ee
where $q$ is the external momentum. For large $k >> q$
the two terms on the rhs have a similar behaviour.
Thus, $\nabla_F$ has an improved
powercounting, and renders the overall logarithmic divergence
in $\bar{\Gamma}(n,m)$ finite, when inserted for $\Delta_F$.
Consequently, we can replace  
$\Delta_F^{[r]}(k+q)$ by $\Delta_F^{[r-j]}(k+q)\;k^2\; \Delta_F^{[j]}(k)$ 
for any $j$, which gives the observed symmetry.

Next, let us see if we can
understand the various transcendentals appearing in the counterterms.
First, we note that $\bar{\Gamma}(3)$ is the first case which delivers
a new topology different from the ladder topology. It is reasonable to expect
that this change in topology is reflected in the overall divergence
$<\bar{\Gamma}(3)>$. And indeed, we find that $<\bar{\Gamma}(3)>$ 
contains $\zeta(3)$, and thus should be related to the trefoil knot
\cite{plb}. 

There are two possible approaches to assign knots to
Feynman diagrams. Either, one maps the possible momentum routings
in Feynman diagrams to link diagrams \cite{habil} or one
investigates the Gauss code of the Feynman diagram under consideration 
\cite{bk15}. Both approaches are intuitive and serve the purpose
to demonstrate the connection between the transcendental numbers
obtained from the Feynman diagrams under
consideration and low-dimensional topology.
Defining these mappings rigorously is beyond the scope of this paper.
A more detailled discussion will be in a forthcoming book \cite{book}.
For our present purposes we are contend to reproduce the pattern which we
observe in Table(1).

Essentially, we want to reproduce the fact that $\bar{\Gamma}(2n+1)$
delivers the $(2,2n+1)$ torus knot (corresponding
to the appearance of $\zeta(2n+1)$ in Table(1))
for $n\geq 1$, and explain
why $\bar{\Gamma}(2n)$ only contains $\zeta(2n-1)$ and relates to
the $(2,2n-1)$ torus knots, $n\geq 2$.

Let us use a method inspired by Gauss codes \cite{bk15,Lou}.
In Fig.(5) we indicate how to obtain the codes
$\{1,1,2,2,\ldots,n,n\}$ and $\{1,2,2,\ldots,n,n,1\}$
for $\bar{\Gamma}(n)$. These are the two distinguished possibilities,
referring to the fact that we can either start reading the code 
between two self-energy insertions, or inside such a bubble.
Further, whenever we go through the graph we have two possibilities
when we come through a subdivergence, passing along one of its
two propagators.

Let us make such a choice for each one-loop
insertion in a string of $r$ subdivergence and let us map this
to a curve with $r$ curls as indicated in Fig.(5), for each such choice.
Fig.(5) gives the example of three one-loop insertions, and an arbitrarily chosen
paths through them is indicated.
For the string $\{1,1,\ldots,n,n\}$ we simply close the curves at both ends,
marked as $a$ and $b$ in the figure.
Running through this curve reproduces the same code as obtained from the 
diagram.
The curve is a circle with some curl in it, but unknotted, for all $n$. 
We thus expect to see rational numbers in the results for
$\bar{\Gamma}(n)$ for all $n$.

Next consider the string $\{1,2,2,\ldots,n,n,1\}$.
We know already how to map the substring $2,2,\ldots,n,n$ to a 
piece of a curve (this is similar to the previous case), but we now simply
define the mapping of  $\{1,2,2,\ldots,n,n,1\}$ to a closed curve
as indicated at the bottom of  Fig.(5). 
We effectively double the strand which runs from $a$ to $b$
on the side which has lesser curl and 
close the curve by once more identifying the points
$a$ and $b$, and the point marked $x$. Each self-energy
which is met by the doubled strand produces two extra crossings. 
Still, passing along
the curve and notating the number of each crossing in the order 
of appearance reproduces
the code. This way of doubling a strand is known as a cabling operation
in knot theory. 
This is in accord with an analysis using momentum routings, and with
the experience gained in \cite{habil}.

Let us count. We need at least three subdivergences for the first knot
to appear, and only addition of two further subdivergences through which
we pass in different manner will generate a knot with two more crossings.
The two more subdivergences are needed because we double the strand
at the side with lesser curl. So we need to generate more curl on each side to
increase the number of crossings in the knot. 
This is in accordance with the transcendentals found in $\bar{\Gamma}(n)$
for all $n$. In this manner, by distributing the amount of curl
on both sides in all posible ways, one generates all transcendentals
observed in Table(1).

Products of $\zeta$'s can be obtained by lifting the
factorization properties of the propagator Eq.(\ref{efac}) to 
knot diagrams, which produces the factor knots corresponding 
to the products of lower lying transcendentals in Table(1).
These expectations are indeed confirmed by explicit
calculations, $\overline{\Gamma}(5)$ in Table(\ref{t1})
provides an example for the appearance of $\zeta(5)$ 
and $\zeta(3)$ in a six-loop example,
$\overline{\Gamma}(7)$ provides $\zeta(7), \zeta(5)$
and $\zeta(3)$ and $\overline{\Gamma}(10)$
provides $\zeta(9), \zeta(7), \zeta(5)$ and $\zeta(3)$,
as well as the expected products of $\zeta$'s and rational contributions.
\begin{figure}
\begin{center}
\epsfysize=14cm \fbox{\epsfbox{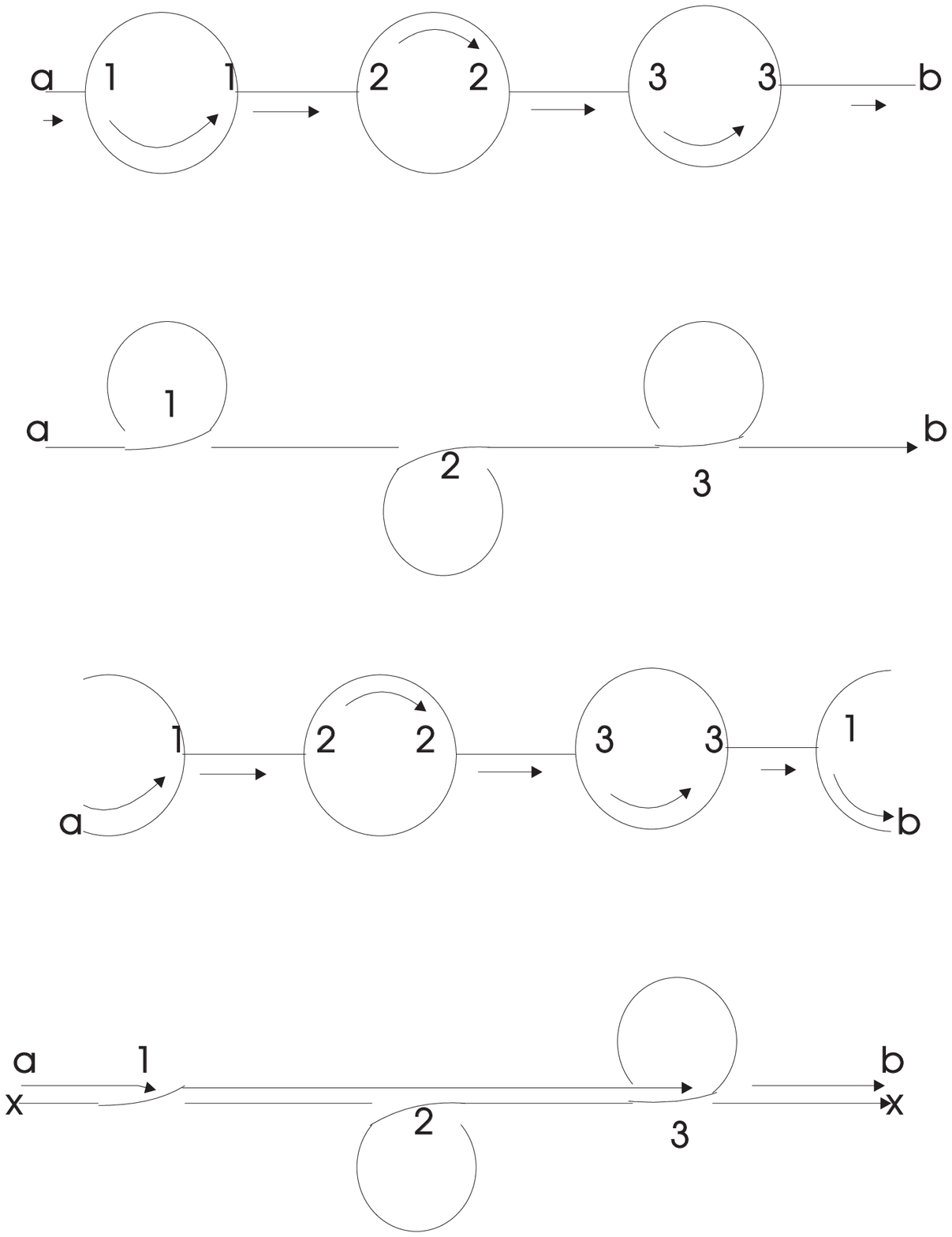}}
\caption{The first line gives a string with three subdivergences.
Running through the graph produces a code. This code is obtained
from the curve below when we run from $a$ to $b$. 
There are two principal choices how to run through a one-loop
subdivergences, which results in curl on different sides.
When we start running in the middle of a subdivergence
(third line) we will assign to this a curve which has a doubled strand
(fourth line).
We double on the side where we have lesser curl, to minimize the
number of extra crossings. In the example considered here this
is irrelevant as we have one curl on each side.
The resulting knot is always a $(2,q)$ torus knot
which matches the transcendentals in Table(1).}
\label{gfcn1}
\end{center}
\end{figure}
There are hints that the presence of even $\zeta$'s ($\zeta(2n)$,
$n>1$, $\sim \pi^{2n}$) is related to the differences in writhe between
the different strands which participate in the cabling operation.
This is currently under investigation.
\section{Continuation to a dressed two-loop graph}
\begin{figure}
\begin{center}
\epsfysize=5cm \fbox{\epsfbox{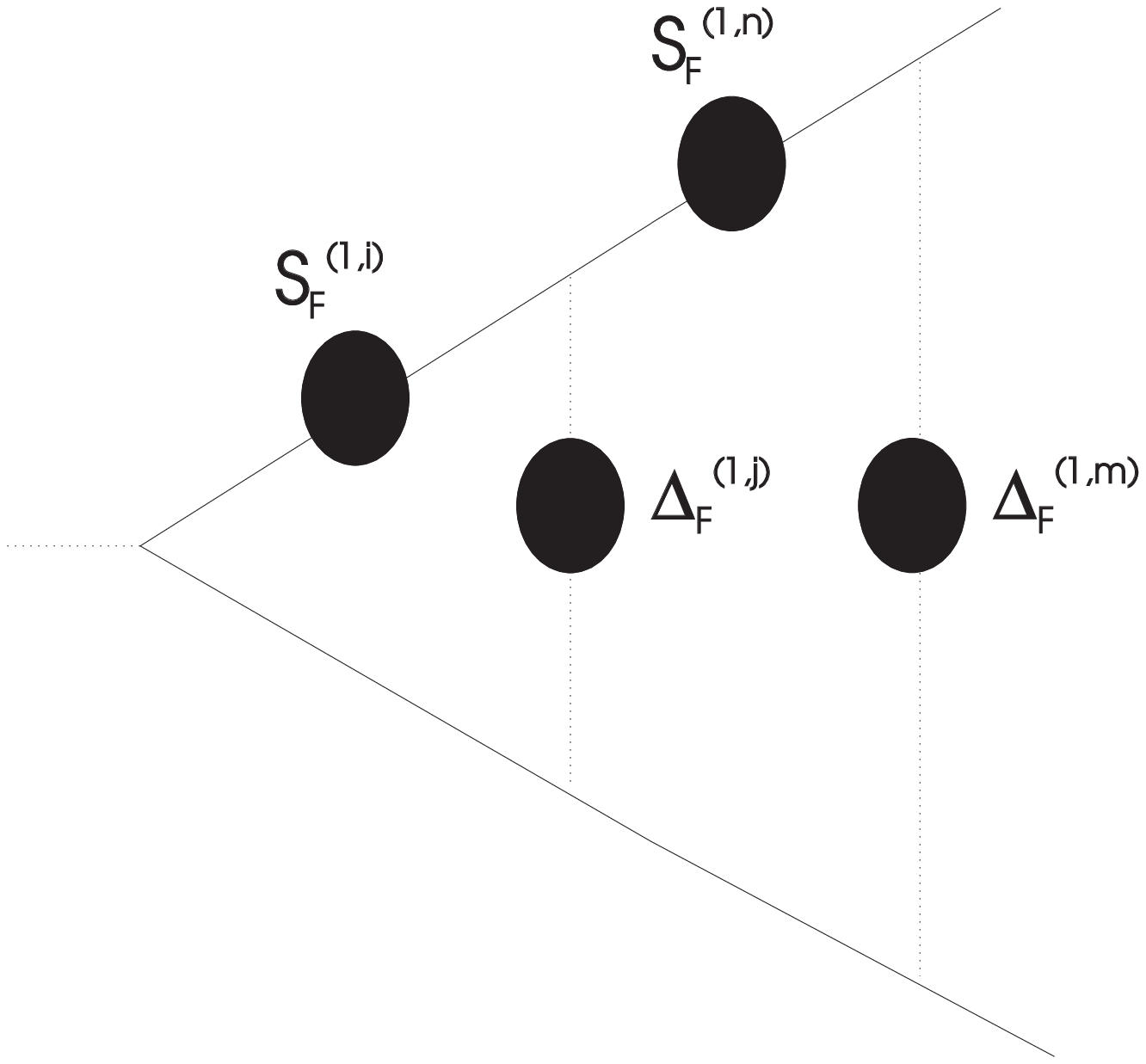}}
\caption{$\bar{\Gamma}(i,j,m,n)$.}
\label{gfc7}
\end{center}
\end{figure}

Our generic topology is indicated in Fig.(\ref{gfc7}), which
we notate by $\overline{\Gamma}(i,j,m,n)$
for a two loop ladder graph with $(i+j+m+n)$ one-loop
subdivergences at the
indicated places. 
It is defined by the following expression (this time we omit
to mention the case without renormalized subdivergences at all -
we know already that all the nice properties will only turn up
after renormalization of subgraphs).
\be
\bar{\Gamma}(i,j,n,m)(q) 
:= 
\int d^Dk \bar{\Gamma}(i,j)(k)\;\frac{1}{\kslash}\;
{\bf S_F}^{[1,n]}(k){\bf \Delta_F}^{[1,m]}(k+q).
\ee
In Table(\ref{t2}) we collect results. We spot some remarkable
properties which, again, are visible only after
elimination of subdivergences.
\begin{table}
\rule{\textwidth}{.5mm}\\
$\overline{\Gamma}(2,0,0,0) := \;\epsilon^{(-1)} (1/2 \;\zeta(3) - 5/12) - 1/12 \;\epsilon^{(-2)} 
+ 1/4 \;\epsilon^{
(-3)} - 1/12 \;\epsilon^{(-4)}$\\
\rule{\textwidth}{.5mm}\\
$\overline{\Gamma}(1,1,0,0) :=  - 5/12 \;\epsilon^{(-1)} - 1/12 \;\epsilon^{(-2)} + 1/4 \;\epsilon^{(-3)} 
- 1/12 
\;\epsilon^{(-4)}$\\
\rule{\textwidth}{.5mm}\\
$\overline{\Gamma}(0,2,0,0) := \;\epsilon^{(-1)} (1/2 \;\zeta(3) - 5/12) - 1/12 \;\epsilon^{(-2)} 
+ 1/4 \;\epsilon^{
(-3)} - 1/12 \;\epsilon^{(-4)}$\\
\rule{\textwidth}{.5mm}\\
$\overline{\Gamma}(2,0,0,1) := \;\epsilon^{(-1)} ( - 3/20 \;\zeta(4) - 1/10 \;\zeta(3)) 
+ \;\epsilon^{(-2)} ( -
 1/10 \;\zeta(3) + 19/60) - 11/60 \;\epsilon^{(-4)} + 1/15 \;\epsilon^{(-5)}$\\
\rule{\textwidth}{.5mm}\\
$\overline{\Gamma}(1,1,0,1) := 19/60 \;\epsilon^{(-2)} - 11/60 \;\epsilon^{(-4)} 
+ 1/15 \;\epsilon^{(-5)}$\\
\rule{\textwidth}{.5mm}\\
$\overline{\Gamma}(0,2,0,1) := \;\epsilon^{(-1)} ( - 3/20 \;\zeta(4) - 1/10 \;\zeta(3)) 
+ \;\epsilon^{(-2)} ( -
 1/10 \;\zeta(3) + 19/60) - 11/60 \;\epsilon^{(-4)} + 1/15 \;\epsilon^{(-5)}$\\
\rule{\textwidth}{.5mm}

$\overline{\Gamma}(2,0,0,2) := \;\epsilon^{(-1)} 
(3/10 \;\zeta(4) - 1/5 \;\zeta(3) + 11/30) + \;\epsilon^{(-2)} (1/5 
\;\zeta(3)
 - 7/45) - 19/90 \;\epsilon^{(-3)} + 
1/45 \;\epsilon^{(-4)} + 13/90 \;\epsilon^{(-5)} - 1/18 \;\epsilon^{(-6)}$\\
\rule{\textwidth}{.5mm}
$\overline{\Gamma}(1,1,0,2) := \;\epsilon^{(-1)} 
( - 7/30 \;\zeta(5) + 1/4 \;\zeta(4) - 11/30 \;\zeta(3) + 11/30) + 
\;\epsilon
^{(-2)} 
( - 1/20 \;\zeta(4) + 1/6 \;\zeta(3) - 7/45) + \;\epsilon^{(-3)} 
( - 1/30 \;\zeta(3) - 19/90) +
 1/45 \;\epsilon^{(-4)} + 13/90 \;\epsilon^{(-5)} - 1/18 \;\epsilon^{(-6)}$\\
\rule{\textwidth}{.5mm}
$\overline{\Gamma}(2,0,1,1) := 
\;\epsilon^{(-1)} (3/10 \;\zeta(4) - 1/5 \;\zeta(3) + 11/30) + \;\epsilon^{(-2)}
 (1/5 
\;\zeta(3)
 - 7/45) - 19/90 \;\epsilon^{(-3)} + 1/45 \;\epsilon^{(-4)} + 13/90 \;\epsilon^{(-5)} - 
1/18 \;\epsilon^{(-6)}$\\
\rule{\textwidth}{.5mm}
$\overline{\Gamma}(1,1,1,1) := 
\;\epsilon^{(-1)} ( - 7/30 \;\zeta(5) + 1/4 \;\zeta(4) - 11/30 \;\zeta(3) + 11/30) + 
\;\epsilon^{(-2)} ( - 1/20 \;\zeta(4) + 1/6 \;\zeta(3) - 7/45) + \;\epsilon^{(-3)}
 ( - 1/30 \;\zeta(3) - 19/90) +
 1/45 \;\epsilon^{(-4)} + 13/90 \;\epsilon^{(-5)} - 1/18 \;\epsilon^{(-6)}$\\
\rule{\textwidth}{.5mm}

$\overline{\Gamma}(3,0,0,0) := \;\epsilon^{(-1)} 
( - 3/5 \;\zeta(4) + 1/5 \;\zeta(3) + 3/10) + \;\epsilon^{(-2)} (1/10 
\;\zeta(
3) + 1/20) + 1/20 \;\epsilon^{(-3)} - 3/20 \;\epsilon^{(-4)} + 1/20 
\;\epsilon^{(-5)}$\\
\rule{\textwidth}{.5mm}
$\overline{\Gamma}(2,1,0,0) := \;\epsilon^{(-1)} 
( - 3/10 \;\zeta(4) - 2/5 \;\zeta(3) + 3/10) + \;\epsilon^{(-2)} (3/10 
\;\zeta
(3) + 1/20) + 1/20 \;\epsilon^{(-3)} - 3/20 \;\epsilon^{(-4)} + 1/20 
\;\epsilon^{(-5)}$\\
\rule{\textwidth}{.5mm}
$\overline{\Gamma}(4,0,0,0) := 
\;\epsilon^{(-1)} (\;\zeta(5) - 3/10 \;\zeta(4) - 7/15 \;\zeta(3) - 7/30) 
+ \;\epsilon^{(-2)} ( 
- 1/10 \;\zeta(4) + 1/5 \;\zeta(3) - 1/30) + \;\epsilon^{(-3)} 
( - 1/15 \;\zeta(3) - 1/30) - 1/30 
\;\epsilon^{(-4)} + 1/10 \;\epsilon^{(-5)} - 1/30 \;\epsilon^{(-6)}$\\
\rule{\textwidth}{.5mm}
$\overline{\Gamma}(3,1,0,0) := \;\epsilon^{(-1)}
 (3/10 \;\zeta(5) + 3/20 \;\zeta(4) - 17/30 \;\zeta(3) - 7/30) + 
\;\epsilon^{(-2)}
 ( - 1/4 \;\zeta(4) + 1/2 \;\zeta(3) - 1/30) 
+ \;\epsilon^{(-3)} ( - 1/6 \;\zeta(3) - 1/30) - 1/30 
\;\epsilon^{(-4)} + 1/10 \;\epsilon^{(-5)} - 1/30 \;\epsilon^{(-6)}$\\
\rule{\textwidth}{.5mm}

$\overline{\Gamma}(3,0,0,1) := \;\epsilon^{(-1)}
 (3/20 \;\zeta(5) - 1/10 \;\zeta(4) - 4/15 \;\zeta(3) - 1/12) + \;\epsilon^{(-2)}
 (7/40 \;\zeta(4) + 11/60 \;\zeta(3) - 2/15) + \;\epsilon^{(-3)} 
( - 2/15 \;\zeta(3) - 1/12) - 1/120 \;\epsilon^{(-4)} + 7/60 \;\epsilon^{(-5)}
 - 1/24 \;\epsilon^{(-6)}$\\
\rule{\textwidth}{.5mm}
$\overline{\Gamma}(2,1,0,1) := 
\;\epsilon^{(-1)} (7/20 \;\zeta(5) + 1/5 \;\zeta(3) - 1/12) + \;\epsilon^{(-2)} (3/40 
\;\zeta(4)
 + 1/4 \;\zeta(3) - 2/15) + \;\epsilon^{(-3)} ( - 1/5 \;\zeta(3) - 1/12) - 
1/120 \;\epsilon^{(-4)} + 7/60
 \;\epsilon^{(-5)} - 1/24 \;\epsilon^{(-6)}$\\
\rule{\textwidth}{.5mm}
$\overline{\Gamma}(4,0,0,1) := 
\;\epsilon^{(-1)} ( - 2/7 \;\zeta(6) - 3/35 \;\zeta(5) + 17/70 \;\zeta(4) - 6/35 
\;\zeta(3)^{2} - 
1/105 \;\zeta(3) + 1/105) + \;\epsilon^{(-2)}
 ( - 8/35 \;\zeta(5) - 4/35 \;\zeta(4) + 38/105
 \;\zeta(3) + 16/105) + \;\epsilon^{(-3)}
 (9/70 \;\zeta(4) - 29/105 \;\zeta(3) + 1/105) + \;\epsilon^{(-4)} (
3/35 \;\zeta(3) + 11/210) + 1/105 \;\epsilon^{(-5)}
 - 17/210 \;\epsilon^{(-6)} + 1/35 \;\epsilon^{(-7)}$\\
\rule{\textwidth}{.5mm}
$\overline{\Gamma}(3,1,0,1) := \;\epsilon^{(-1)}
 ( - 5/14 \;\zeta(6) + 39/70 \;\zeta(5) - 1/28 \;\zeta(4) + 29/70 
\;\zeta
(3)^{2}
 + 67/210 \;\zeta(3) + 1/105) + \;\epsilon^{(-2)}
 ( - 3/14 \;\zeta(5) - 31/140 \;\zeta(4) + 
37/210 \;\zeta(3) + 16/105) + \;\epsilon^{(-3)}
 (27/140 \;\zeta(4) - 73/210 \;\zeta(3) + 1/105) + 
\;\epsilon^{(-4)} 
(9/70 \;\zeta(3) + 11/210) + 1/105 \;\epsilon^{(-5)} - 17/210 \;\epsilon^{(-6)}
 + 1/35 \;\epsilon^{(-7)}$\\
\rule{\textwidth}{.5mm}

\caption{$<\overline{\Gamma}(i,j,m,n)>$}
\label{t2}
\end{table}

\begin{figure}
\begin{center}
\epsfysize=8cm \fbox{\epsfbox{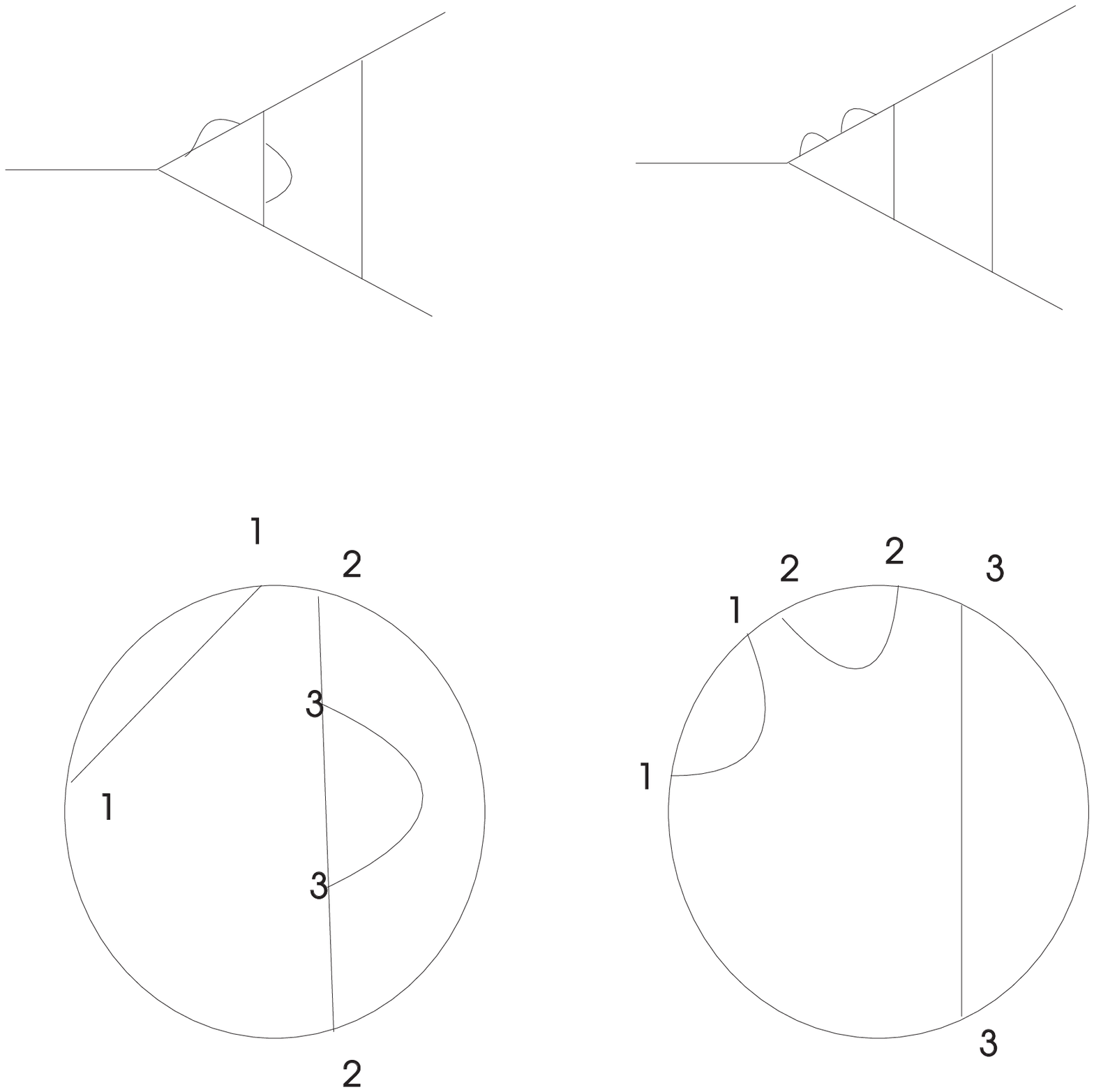}}
\caption{$<\bar{\Gamma}(1,1,0,0)>$, $<\bar{\Gamma}(2,0,0,0)>$
and their topologies and codes drawn below.
$\bar{\Gamma}(1,1,0,0)$ reproduces the ladder topology
$\{1,2,3,3,2,1\}$ and $\bar{\Gamma}(2,0,0,0)$ reproduces the same topology
as $\bar{\Gamma}(3)$: $\{1,1,2,2,3,3\}$.}
\label{gfcn2}
\end{center}
\end{figure}

The case $<\overline{\Gamma}(2,0,0,0)>$ delivers the same value
as $<\overline{\Gamma}(0,2,0,0)>$. It contains $\zeta(3)$.
The symmetry is obvious when studying the
analytic expressions for the two cases, from the fact that
$G(i_1,i_2)=G(i_2,i_1)$. As link diagrams, it is obvious
that any link diagram for  $\overline{\Gamma}(2,0,0,0)$
is a link diagram for $\overline{\Gamma}(0,2,0,0)$ as well.
Forgetting the points where external particles couple, the
diagrams are topologically equivalent.
In contrast, $<\overline{\Gamma}(1,1,0,0)>$ is free of $\zeta(3)$.
This is clear from the fact that it is topologically equivalent
to a ladder diagram, see Fig.(\ref{gfcn2}), while $<\overline{\Gamma}(2,0,0,0)>$
is topologically equivalent to $<\overline{\Gamma}(3)>$. 
Fig.(\ref{gfcn2}) demonstrates this clearly.
Now a Gauss code analysis for a $n$-loop ladder diagram reveals that
all its Gauss codes are equivalent to the code $\{1,1,\ldots,n,n\}$, which
explains its rationality in this language. 

This points towards  an even more striking fact hidden in
Table(\ref{t2}). While
$<\overline{\Gamma}(2,0,0,0)>=
<\overline{\Gamma}(0,2,0,0)>$ contain knotted -transcendental-
and unknotted -rational- codes, $<\overline{\Gamma}(1,1,0,0)>$
contains only the rational code.  
This is in striking agreement with the fact that the
rational contribution for the $(2,0,0,0)$ and $(0,2,0,0)$
cases is {\bf the same} as the $(1,1,0,0)$ contribution,
cf.~Table(2).

To test these phenomena, we calculated further examples.
Some of them are collected in Table(2).
By inspection, we indeed see that the rational parts
of\\ 
$\{\overline{\Gamma}(2,0,1,1),\overline{\Gamma}(1,1,1,1)\}$,\\
$\{\overline{\Gamma}(2,0,0,2),\overline{\Gamma}(1,1,0,2)\}$,\\
$\{\overline{\Gamma}(3,0,0,0),\overline{\Gamma}(2,1,0,0)\}$,\\
$\{\overline{\Gamma}(3,0,0,1),\overline{\Gamma}(2,1,0,1)\}$,\\
$\{\overline{\Gamma}(4,0,0,0),\overline{\Gamma}(3,1,0,0)\}$,\\
$\{\overline{\Gamma}(4,0,0,1),\overline{\Gamma}(3,1,0,1)\}$\\
are indeed identical, in perfect agreement with the analysis 
of Gauss codes or link diagrams.\footnote{We also have the symmetry
$(i,j,m,n)=(i,j,m^\prime,n^\prime)$, $m+n=m^\prime+n^\prime$,
which one can observe in Table(2). It is similar to the symmetries of
$\bar{\Gamma}(n)$.}
Again, Table(2) exhibits typical
examples, while the reported phenomenon was confirmed 
systematically to higher orders with results
which are too long to be reproduced here.

This indicates new relations between Feynman diagrams,
not used and explored so far. Note that these relations are
hidden in various ways.
First of all, trivial prefactors resulting from various
spin factors of the particles hide these relations.
For this reason we normalized our graphs omitting the factors
in eqs.(\ref{ese},\ref{epe}), as mentioned before. These factors seem to
play the same role as group theoretic factors in
Chern Simons theory, if one likes this analogy,
while the knottishness - the topology- is in the
transcendentals.

Second, such relations are impossible to observe as long as 
one does not restrict oneself to the consideration
of overall divergences, the leading symbol of the graph,
so to speak, by elimination of subdivergences.
Again, all symmetries and number theoretic properties are only
apparent after proper subtraction of subdivergences, 
emphasizing the importance of studying proper overall divergences,
to see the connection with link theory.

In the next section we mention two further examples, as they are instructive
for the reader.
\section{Higher order dressing}
\begin{figure}
\begin{center}
\epsfysize=7cm \fbox{\epsfbox{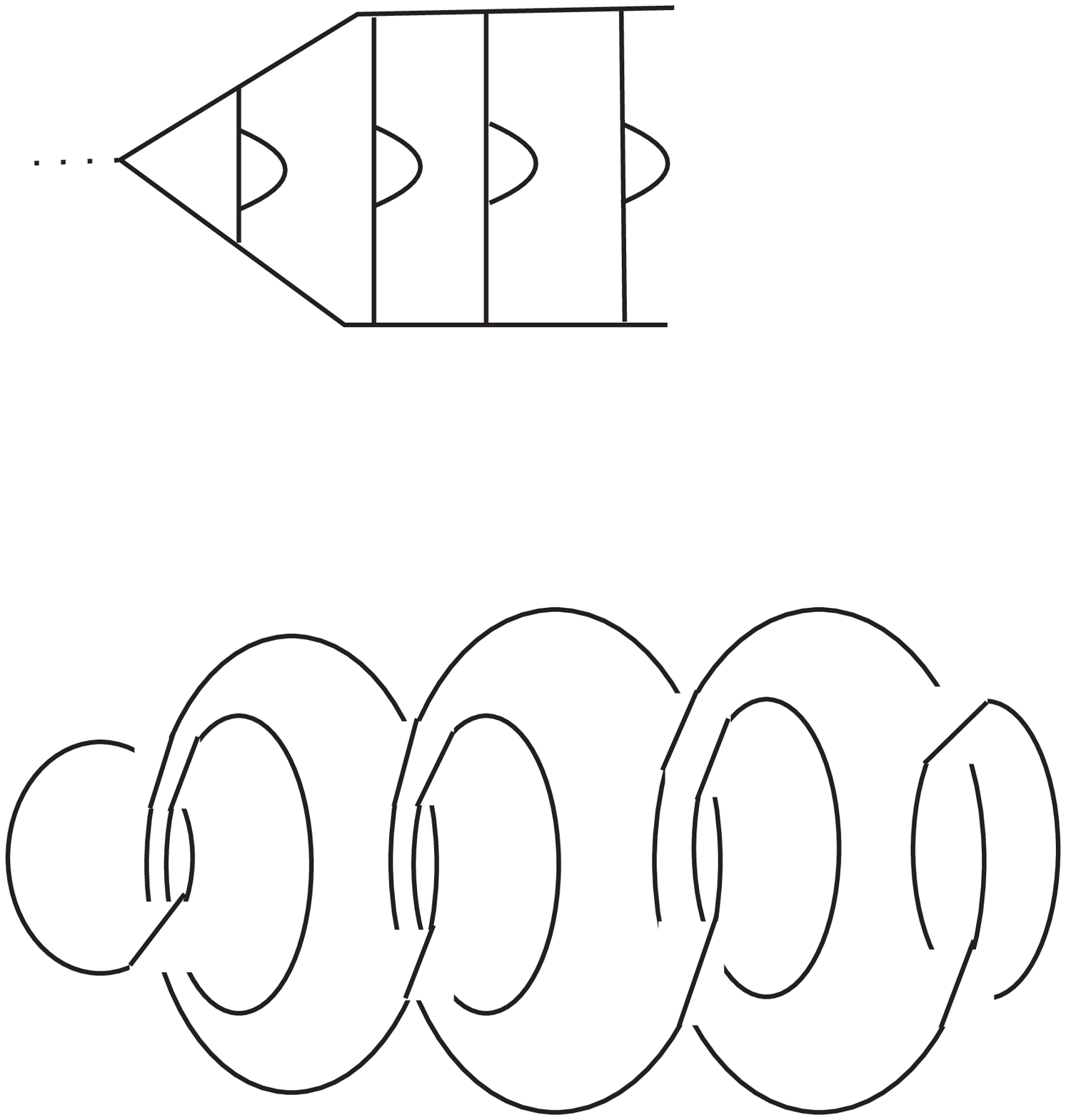}}
\caption{$\overline{\Gamma}_n(1,0)$. As a Gauss code is not
available in this situation, we indicate the entanglement of
a link diagram associated to the graph.
The 
link diagram is knot free.}
\label{gfc9}
\end{center}
\end{figure}

We can generalize the examples of the previous section
when we increase the number of loops in the ladder,
but still have chains of one-loop subdivergences at internal
lines. A particularly interesting example is the function
$\overline{\Gamma}_n(1,0)$, defined in
Fig.(\ref{gfc9}). The figure also explains the rationality
of $\overline{\Gamma}_n(1,0)$. We refer to the methods of \cite{habil,plb} 
in this
case, as a Gauss code is not available.\footnote{This is due to the fact
that there is no closed non-selfintersecting curve which runs through all
the vertices of the diagram.} 
This rationality agrees with its evaluation in terms of $G$-functions:
\begin{eqnarray}
\overline{\Gamma}_2(1,0) & = & 
- 1/3 \;\epsilon^{(-1)} - 5/24 \;\epsilon^{(-2)} 
+ 1/3 \;\epsilon^{(-3)} - 1/8 \;\epsilon^{(-4)},\\
\overline{\Gamma}_3(1,0) & = & 
4/15 \;\epsilon^{(-1)} + 13/30 \;\epsilon^{(-2)} - 
17/240 \;\epsilon^{(-3)} - 43/240 \;\epsilon^{(-4)}\nonumber\\ 
 & & + 5/48 \;\epsilon^{(-5)} - 1/48 \;\epsilon^{(-6)}.
\end{eqnarray}
We omit to give results for higher loop orders, but again
calculations confirm rationality for all cases which have been
tested.\footnote{Usually, 
memory and CPU time restrictions allowed for tests
up to 10-20 loops.}
The cancellation of transcendentals is highly non-trivial,
due to the presense of multiple counterterms generated
by the subdivergences. 

\begin{figure}
\begin{center}
\epsfysize=7cm \fbox{\epsfbox{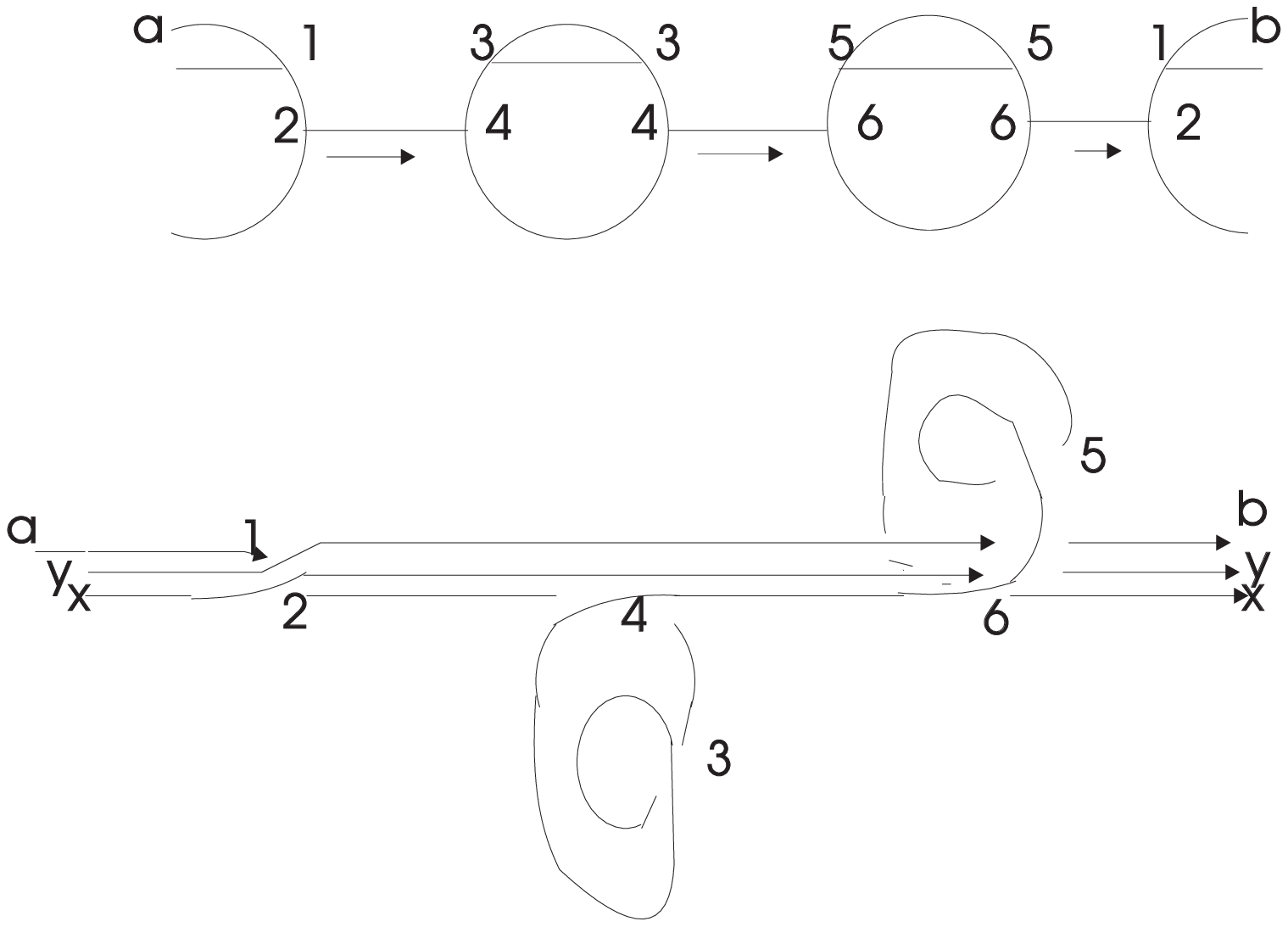}}
\caption{One further cabling of the strand from $a$ to $b$ 
produces a knot whose presence matches the transcendentals in Table(3),
here demonstrated for $\bar{\Gamma_2}(3)$. The knot is obtained by
the identifications at $x,y$ and $a$ with $b$, and turns out to be
the $(2,5)$ torus knot.}
\label{gfc10}
\end{center}
\end{figure}

Further, Fig.(\ref{gfc10}) considers dressings with two-loop rainbows,
and shows the appearance of  knots matching the transcendentals
in the overall divergence, given in Table(\ref{t3}).
\begin{table}
\rule{\textwidth}{.5mm}\\
$\Gamma_2(1) := 2/3\;\epsilon^{(-1)} - 1/2\;\epsilon^{(-2)} + 
1/6\;\epsilon^{(-3)}$\\
\rule{\textwidth}{.5mm}\\
$\Gamma_2(2) := \;\epsilon^{(-1)}(1/80\;\zeta(4) + 4/15) - 2/15\;\epsilon^{(-2)}
 + 17/60\;\epsilon^{(-3)} - 1/5\;\epsilon^{(-4)} 
+ 1/20\;\epsilon^{(-5)}$\\
\rule{\textwidth}{.5mm}\\
$\Gamma_2(3) := \;\epsilon^{(-1)}( - 47/336\;\zeta(6) + 11/56\;\zeta(5) + 3/140
\;\zeta(4) + 
13/70\;\zeta(3)^{2} + 
9/70\;\zeta(3) + 2/7) + \;\epsilon^{(-2)}( - 11/280\;\zeta(5) - 3/28\;\zeta(4) + 
1/70\;\zeta(3) 
+ 1/35) + \;\epsilon^{(-3)}(3/140\;\zeta(4) - 1/14\;\zeta(3) + 1/14) + 
\;\epsilon^{(-4)}
(1/70\;\zeta(3) - 39/280) + 47/280\;\epsilon^{(-5)} - 5/56\;\epsilon^{(-6)} + 
1/56\;\epsilon^{(-7)}$\\
\rule{\textwidth}{.5mm}\\

$\overline{\Gamma}_2(4) := 
\;\epsilon^{(-1)} ( - 
575/1152 \;\zeta(8) + 167/168 \;\zeta(7) - 137/216 \;\zeta(6) + 18/35 
\;\zeta(5) \;\zeta(3) + 18/35 \;\zeta(5) + 13/140 \;\zeta(4)^{2} 
- 26/35 \;\zeta(4) \;\zeta(3) - 17/84
 \;\zeta(4) + 286/315 \;\zeta(3)^{2} + 
2/315 \;\zeta(3) + 136/315) + \;\epsilon^{(-2)} ( - 167/1008 
\;\zeta(7) + 113/1008 \;\zeta(6) - 6/35 \;\zeta(5) + 
13/105 \;\zeta(4) \;\zeta(3) - 19/210 \;\zeta(4)
 - 26/105 \;\zeta(3)^{2} - 
17/126 \;\zeta(3) + 4/45) + \;\epsilon^{(-3)} ( - 113/6048 \;\zeta(6) + 
2/15 \;\zeta(4) + 13/315 \;\zeta(3)^{2} - 19/315 \;\zeta(3) + 2/63) + 
\;\epsilon^{(-4)} ( - 1/14 \;\zeta(4)
 + 4/45 \;\zeta(3) - 1/63) + \;\epsilon^{(-5)} (1/84 \;\zeta(4) 
- 1/21 \;\zeta(3) + 377/5040) 
+ \;\epsilon^{(-6)} (1/126 \;\zeta(3) - 59/504) + 
25/252 \;\epsilon^{(-7)} - 
1/24 \;\epsilon ^{(-8)} + 1/144 \;\epsilon^{(-9)}$\\
\rule{\textwidth}{.5mm}

$\overline{\Gamma}_2(5) := 
\;\epsilon^{(-1)} 
( - 5199/3520 \;\zeta(10) + 11333/3168 \;\zeta(9) - 45701/12672 \;\zeta(8) 
+ 565/616 \;\zeta(7) \;\zeta(3) + 
28505/11088 \;\zeta(7) + 
25/56 \;\zeta(6) \;\zeta(4) - 25/12 
\;\zeta(6) \;\zeta(3) - 
65/63 \;\zeta(6) + 43/88 \;\zeta(5)^{2} 
- 133/88 \;\zeta(5) \;\zeta(4) + 860/231 \;\zeta(5) \;\zeta(3) 
+ 631/1848 \;\zeta(5) + 53/77 \;\zeta(4)^{2}
 - 135/616 \;\zeta(4) \;\zeta(3)^{2} 
- 1667/924 \;\zeta(4) \;\zeta(3) - 47/168 \;\zeta(4) + 15/44 \;\zeta(3)^{3} 
+ 6617/5544 
\;\zeta(3)^{2} - 485/1386 \;\zeta(3) + 538/693) + \;\epsilon^{(-2)}
 ( - 1619/3168 \;\zeta(9) + 2737/2816 \;\zeta(8) -
 1835/1584 \;\zeta(7) + 25/84 \;\zeta(6) \;\zeta(3) + 25/63 \;\zeta(6) + 19/88 
\;\zeta(5) \;\zeta(4) - 133/132 \;\zeta(5) \;\zeta(3) - 
701/1848 \;\zeta(5) - 69/352 \;\zeta(4)^{2} + 
212/231 \;\zeta(4) \;\zeta(3) + 15/1232 \;\zeta(4) 
- 15/308 \;\zeta(3)^{3} - 1667/2772 \;\zeta(3)^{2} 
- 47/252 \;\zeta(3) + 19/99) + \;\epsilon^{(-3)} ( - 391/2816 \;\zeta(8) 
+ 305/1056 \;\zeta(7) 
- 25/252 \;\zeta(6) + 19/132 \;\zeta(5) \;\zeta(3) + 3/44 \;\zeta(5) + 
69/2464 \;\zeta(4)^{2} - 23/88 \;\zeta(4) \;\zeta(3) + 
185/3696 \;\zeta(4) + 212/693 \;\zeta(3)^{2} + 5/616 \;\zeta(3) + 1/18
) + \;\epsilon^{(-4)} ( - 305/7392 \;\zeta(7) + 1/33 \;\zeta(5) + 
23/616 \;\zeta(4) \;\zeta(3) - 251/1848 \;\zeta(4) - 23/264 \;\zeta(3)^{2} 
+ 185/5544 \;\zeta(3) + 37/2772) + \;\epsilon^{(-5)} 
( - 7/264 \;\zeta(5) + 97/924 \;\zeta(4) + 23/1848 \;\zeta(3)^{2} 
- 251/2772 \;\zeta(3) + 
1/72) + \;\epsilon^{(-6)} (1/264 \;\zeta(5) - 7/176 \;\zeta(4) + 97/1386 \;\zeta(3) - 
263/7392) + \;\epsilon^{(-7)} (1/176 \;\zeta(4) - 7/264 \;\zeta(3) + 1699/22176) + 
\;\epsilon^{(-8)} (1/264 \;\zeta(3) - 139/1584) + 91/1584 \;\epsilon^{(-9)} - 
7/352 \;\epsilon ^{(-10)} + 1/352 \;\epsilon^{(-11)}$\\
\rule{\textwidth}{.5mm}

\caption{Two-loop rainbow dressings.}
\label{t3}
\end{table}
Note that $\overline{\Gamma_2}(2)$ does not provide any knot-number,
but delivers the transcendental $\zeta(4)$, indicating some extra
writhe in the diagrams. Further, an analysis with the help of Gauss codes
suggest that $\overline{\Gamma_2}(2)$ should contain
$\zeta(5)$, while $\overline{\Gamma_2}(3)$ and
$\overline{\Gamma_2}(4)$ should have $\zeta(7)$ and $\zeta(9)$,
respectively.
Which is indeed the case.   

\section{Other field theories}
One can extend the results considered here to other renormalizable
theories. For clarity, we presented our examples for the case
of a massless Yukawa theory.
To incorporate other theories, one
can resort to the matrix calculus proposed in \cite{habil}. This becomes
necessary due to the different formfactors which are present in general,
resulting from all possible spin structures.
One eventually ends up with similar results. One should not forget that one
has to normalize Green functions in a manner similar to what we did
for Yukawa theory.
Table(\ref{t4}) gives the results 
of  Table(\ref{t1}) for the QED case. 
Considering the vertex correction at zmt, we are confronted with
$\overline{\Gamma}_\mu^{[1]}(i)$, which is defined in the obvious manner.
We then allow for a varying number of $i_1$ one-loop subdivergences
in the fermion line and $i_2$ subdivergences in the photon line, $i=i_1+i_2$.
A moment of thinking ensures that
part of the problem reduces to the concatenation of $G$-functions
that we have already considered for Yukawa theory. 
But QED is much more demanding:
inserting the $k_\mu k_\nu$ part of the boson propagator,
one is also confronted with $G$ functions which have intermediate
IR divergences, as for example
\be
\int d^Dk \frac{1}{[k^2]^2[(k+q)^2]^{j\epsilon}}.
\ee
In Table(\ref{t4}) we only show the results for the
part of the vertex which was infected by such $G$-functions,
and omit the contributions which were similar to the Yukawa case
from the start.
Again counterterms are subtracted in the $\overline{MS}$-scheme. 
\begin{table}
\rule{\textwidth}{.5mm}\\
$\overline{\Gamma}_{QED}(1) := 4 \;\epsilon^{(-1)} - 3 \;\epsilon^{(-2)}$\\
\rule{\textwidth}{.5mm}\\
$\overline{\Gamma}_{QED}(2) :=  - 4/3 \;\epsilon^{(-1)} - 8/3 \;\epsilon^{(-2)} + 2 \;\epsilon^{(-3)}$\\
\rule{\textwidth}{.5mm}\\
$\overline{\Gamma}_{QED}(3) := \;\epsilon^{(-1)} ( - 3 \;\zeta(3) + 1) + \;\epsilon^{(-2)} 
+ 2 \;\epsilon^{(-3)} - 3/2 \;\epsilon^{(-4)}$\\
\rule{\textwidth}{.5mm}\\
$\overline{\Gamma}_{QED}(4) := \;\epsilon^{(-1)} (18/5 \;\zeta(4) - 16/5 \;\zeta(3) - 4/5) + 
\;\epsilon^{(-2)} (12/5 \;\zeta(3) - 4/5) - 4/5 \;\epsilon^{(-3)}
 - 8/5 \;\epsilon^{(-4)} + 6/5 \;\epsilon^{(-5)}$\\
\rule{\textwidth}{.5mm}\\

\caption{Two- to five-loop QED examples.}
\label{t4}
\end{table}
We see that the same pattern arises as before.
No transcendentals up to three loops, and then $\zeta(3)$ plus
rational contributions at four loops.

Finally, let us comment on two results obtained by other authors.
In both cases, calculations were pushed to the four-loop
level, while investigating $\beta$-functions of some sort.

First, the investigation of $\phi^4$ theory in \cite{kast}
considers the four-loop level.
The four-loop result  is in precise
agreement with our expectations. Again, $\zeta(3)$ appears at the
four-loop level, from diagrams where we expect it to occur.
In fact, the circle chain integral (as the author calls it)
$I_3^{cc}$ is topologically equivalent to our
four-loop graph in the first example, and thus we expect the
author to find a contribution containing $\zeta(3)$ as well
as a rational part, which is indeed the case.

Second, in \cite{JJN}, it is conjectured that the appearance
of $\zeta(3)$ at only three places out of  twelve
different topologies (cf.~Fig.4 in the paper) could be explained
using knot theory. With the experience from the results presented
here, this is indeed the case. The three cases (Fig.4F,G,H in the
paper)
which deliver
$\zeta(3)$ in the results of \cite{JJN} are again topologically
equivalent 
to the four-loop case in our first
example, while all the other diagrams only provide
link diagrams free of knots, as they should. In our notation, they all
belong to cases similar to $\overline{\Gamma}_n(1,0)$.
\section{Conclusions}
In this paper we compared overall divergences of Feynman graphs
containing subdivergences with link diagrams. 
Using recent results on a connection between knot theory
and renormalization theory, we explored the appearance of transcendentals
in the overall divergences. 
Our results indicate that also graphs which do contain
subdivergences can be
investigated using link and knot theory. This closes the
before-mentioned gap between results 
concerning topologically trivial ladder graphs and results concerning
graphs of complicated topology, but free of subdivergences. In this paper
we considered the middle ground inbetween.

Specifically, we found that disjoint subdivergences
of a simple structure do provide transcendentals in
accordance with link diagrams assigned to them
in the manner proposed in \cite{habil,pisa,bgk,bk15}.
In the most simple case, we dressed a one-loop
skeleton graph with chains of one-loop subdivergences.
Link diagrams for this case suggested the appearance
of all $(2,q)$ torus knots, with the highest $q$ determined
by the loop number. Table(\ref{t1}) summarizes
these results.

Using only very basic properties of link diagrams,
we discovered a new relation between rational parts
of overall counterterms, as it is dramatically exemplified
in Table(\ref{t2}). 
All the Feynman diagrams considered there contain 
knot-free link diagrams, but some contain others 
link diagrams as well.
We found that whenever a knot-free link diagram results
from various different diagrams, the rational
part of their contribution to the overall divergence
is the same. This not only confirms that
knots should be associated with transcendentals, but
establishes new relations between diagrams.
Apart from our approach via link theory, there is no other
explanation for such results available.
It indicates that the resolution of the topology of a Feynman
diagram in terms of different link diagrams is a meaningful tool
to understand the number-theoretic properties of counterterms.
The way we assign knots of increasing complexity to these diagrams is by way
of a cabling operation, which indicates further systematics in this approach
which deserves further study.

These relations between diagrams were not found before, mainly,
one guesses,  because
one barely calculates overall divergent quantities by
doing the renormalization of subdivergences graph by graph.
Multiplicative renormalization screens these new findings.
Only when one renormalizes graph by graph, one can observe
these relations.

It seems that a reorganization
of perturbative results ordered with respect to transcendentals
is favourable. Our results indicate
that the forest structure of a graph has a deep connection
to link theory, where the way how forests are mutually disjoint or
nested is reflected in the entanglement of link diagrams
which one can assign to them. 

Further results confirmed our expectations, and we also 
gave some results for QED. We expect that the patterns observed
here are true for a renormalizable theory in general.\footnote{At least
in an even dimensional field theory. In odd dimensions,
one might find a different set of transcendentals. I thank
D.~Broadhurst for stressing this point.}

While already results
for diagrams without subdivergences 
deliver evidence towards a connection between
the theory of links and knots, number theory, and the problem of divergences
in a pQFT, it is apparent that the full richness
of this new connection arises when one allows for the general
case. 
In this new area, much remains to be done, and the results reported here
are only the first few steps towards an understanding of this
connection.
It still seems that the role which UV divergences of a 
quantum field theory play is not fully explored yet.
Here is not the space to muse about connections to recent
developments in mathematics, and we refer to \cite{book}
for such purposes.

In recent results, \cite{4TR,BK4},
results are obtained clarifying the role of the four-term 
relation in counterterms free of subdivergences. 
These results still suffer from an exclusion of graphs with subdivergences.
In the light of the results here we hope to be able to report
on some progress with the four-term relation concerning graphs
with subdivergences soon.
\section*{Acknowledgements}
Foremost, I have to thank David Broadhurst for the enthusiam
and vigor with which he picked up these new ideas and turned
them into a most fruitful collaboration. With equal pleasure
I thank Bob Delbourgo for uncounted discussions and support.
Also, I like to thank Karl Schilcher and J\"urgen
K\"orner for interest and encouragement, and the DFG for support.

\end{document}